\documentclass[11pt]{article}

\usepackage[T1]{fontenc}
\usepackage[utf8]{inputenc}
\usepackage{lmodern}

\usepackage{geometry}
\usepackage{amsmath,amssymb,amsthm}
\usepackage{physics}
\usepackage{graphicx}
\usepackage{bm}
\usepackage{mathrsfs}

\usepackage{hyperref}
\hypersetup{
  colorlinks=true,
  linkcolor=blue,
  citecolor=blue,
  urlcolor=blue
}

\usepackage{cite}

\theoremstyle{plain}

\theoremstyle{definition}

\title{Gauge-Invariant Gravitational Wave Polarization in Metric $f(R)$ Gravity with Cosmological Implications}

\author{%
\parbox{\textwidth}{%
\centering
Ramesh Radhakrishnan$^{1}$\footnotemark[1], 
David McNutt$^{2}$, 
Delaram Mirfendereski$^{3}$, 
Alejandro Pinero$^{4}$,\\
Eric Davis$^{5}$, 
William Julius$^{6}$, 
Gerald Cleaver$^{7}$
}
}

\date{} % leave empty for arXiv

\begin{document}
\maketitle

\footnotetext[1]{Corresponding author: ramesh\_radhakrishna1@baylor.edu}
\footnotetext[2]{Department of Mathematics, The Royal Norwegian Naval Academy, Bergen, Norway}
\footnotetext[3]{Department of Physics \& Astronomy, University of Texas Rio Grande Valley, Texas, USA}
\footnotetext[4]{Department of Physics \& Astronomy, University of Texas Rio Grande Valley, Texas, USA}
\footnotetext[5]{Department of Physics, SUNY Albany, Albany, NY, USA}
\footnotetext[6]{Department of Physics \& Astronomy, Baylor University, Waco, Texas, USA}
\footnotetext[7]{Department of Physics \& Astronomy, Baylor University, Waco, Texas, USA}

\begin{abstract}
We develop a fully gauge-invariant analysis of gravitational-wave polarizations in metric $f(R)$ gravity with a particular focus on the modified Starobinsky model $f(R)=R+\alpha R^{2}-2\Lambda$, whose constant-curvature solution $R_{d}=4\Lambda$ provides a natural de Sitter background for both early- and late-time cosmology. Linearizing the field equations around this background, we derive the Klein--Gordon equation for the curvature perturbation $\delta R$ and show that the scalar propagating mode acquires a mass $m_{\psi}^{2}=1/(6\alpha)$, highlighting how the same scalar degree of freedom governs inflationary dynamics at high curvature and the propagation of gravitational waves in the current accelerating Universe. Using the scalar--vector--tensor decomposition and a decomposition of the perturbed Ricci tensor, we obtain a set of fully gauge-invariant propagation equations that isolate the contributions of the scalar, vector, and tensor modes in the presence of matter. We find that the tensor sector retains the two transverse--traceless polarizations of General Relativity, while the scalar sector supports a massive breathing/longitudinal mode determined by the massive scalar propagating mode. Through the geodesic deviation equation—computed both in a local Minkowski patch and in fully covariant de Sitter form—we independently recover the same polarization content and identify its tidal signatures. The resulting framework connects the extra scalar polarization to cosmological observables, providing a unified, gauge-invariant link between gravitational-wave phenomenology and the cosmological implications of metric $f(R)$ gravity.
\end{abstract}


\vspace{1em}
%%%%%%%%%%%%%%%%%%%%%%%%%%%%%%%%%%%%%%%%%%

\section{Introduction}\label{section:intro}

The study of gravitational-wave (GW) polarizations provides a powerful way to distinguish General Relativity (GR) from alternative theories of gravity. In the classic classification by Eardley, Lee, and Lightman using the Newman--Penrose (NP) formalism~\cite{Eardley}, a general metric theory of gravity may contain up to six possible GW polarization states.  In GR only two of these, the familiar tensor ``plus'' ($\oplus$) and ``cross'' ($\otimes$) modes, are present, corresponding to the two radiative degrees of freedom (DoF) of the metric.  Extensions of GR often introduce additional scalar and/or vectorial modes whose presence modifies the relative displacement of freely-falling test particles.

A particularly well-known example is Brans--Dicke theory~\cite{Brans_Dicke}, in which the scalar field gives rise to an additional transverse breathing mode. More generally, recent analyses using both the NP formalism and the irreducible
(3+1) decomposition~\cite{Alves_2024,Flanagan_2005} have confirmed that the number of NP polarization states does not necessarily coincide with the number of radiative DoF in a theory.  This mismatch appears naturally in scalar--tensor theories and in metric $f(R)$ gravity, where the Ricci scalar perturbation introduces a new massive scalar propagating mode that obeys a Klein--Gordon equation with an effective mass $m_{\psi}$~\cite{Starobinsky1980,DeFelice2010,Sotiriou2010}. For massless propagation, this scalar mode produces a single independent polarization: a transverse breathing distortion of a ring of test particles. When the scalar mode is massive ($m_{\psi}\neq 0$), however, the scalar sector induces a mixture of transverse breathing and longitudinal motion along the propagation direction; these two NP amplitudes are not independent but jointly encode a single scalar radiative DoF.  Thus, metric $f(R)$ gravity contains three radiative DoF but may exhibit up to four NP polarization amplitudes in the massive case.

It is important to note recent discussions regarding the interpretation of NP quantities in theories containing massive modes. For gravitational waves whose group velocity differs
from the speed of light, some NP components that vanish in GR no longer vanish identically, raising subtleties concerning the mapping between NP scalars and physical polarizations~\cite{Eardley}.
This occurs because nonluminal propagation renders the wave vector non-null, so the standard NP null tetrad cannot be aligned with the direction of propagation, and the usual decoupling between radiative and non-radiative components breaks down ~\cite{Gong_2018,Alves_2024}. These issues do not invalidate the NP approach but motivate complementary gauge-invariant formalisms.

Several modified gravity models exhibit similar features.  In modified Gauss--Bonnet gravity $f(G)$, for example, tensor waves propagate as in GR while an additional massive scalar mode appears~\cite{Inagaki_2020}.  In massive gravity theories studied via Bardeen variables~\cite{Bardeen_variables}, a normally non-radiative scalar mode becomes dynamical, constituting the helicity-0 component of a massive graviton.  By contrast, quadratic theories
such as Einstein--Dilaton--Gauss--Bonnet and dynamical Chern--Simons gravity can display the same polarization content as GR in their linearized limit~\cite{WagleSafferYunes2019}. Extensions involving explicit matter--geometry couplings, including $f(R,T)$ and $f(R,T^{\phi})$ gravity~\cite{Khlopov}, further illustrate the subtlety of polarization counting.  Although $f(R,T)$ and $f(R)$ gravity share the same far-field polarization structure in vacuum ($T=0$), their source-dependent dynamics differ.  The $f(R,T^{\phi})$ theory, involving an additional scalar $\phi$, leads to distinguishable polarization patterns even in vacuum.  These analyses also emphasize that the scalar mode $\psi$ belonging to $f(R)$ gravity must be clearly distinguished from any additional matter scalar $\phi$.

Gravitational radiation in linearized metric $f(R)$ gravity has been studied in both power-series models
\begin{equation}
    f(R)=\sum_{n=0}^{N}a_{n}R^{n},
\end{equation}
and in specific subclasses such as the Starobinsky model.  Solar-system tests show that metric $f(R)$ gravity reproduces light deflection with the same post-Newtonian parameter $\Gamma=1$ as GR, although perihelion precession can differ~\cite{PhysRevD.83.104022}.  Studies of waveform phases in
extreme-mass-ratio inspirals suggest that deviations from GR may be detectable in some regimes.  Polarization analyses performed in both $f(R)$ and Horndeski theories~\cite{Gong_2018} highlight the challenge of detecting longitudinal scalar modes with laser interferometers, whereas Pulsar Timing Arrays may offer greater sensitivity. The mixed longitudinal--breathing nature of the massive scalar propagating mode has been explicitly confirmed using the geodesic deviation equation~\cite{LiangGongHouLiu2017}. Additional applications of $f(R)$ gravity include the study of gravitational
radiation from white dwarfs with sub- and super-Chandrasekhar masses, where all relevant polarization amplitudes were estimated using Green-function methods~\cite{ThorneKovacs1975}.

A de Sitter background is particularly well motivated for analyzing the propagation of tensor and scalar modes in metric $f(R)$ gravity. It provides an excellent approximation to late-time cosmic acceleration driven by dark energy and also captures the quasi-exponential ``slow-roll'' inflationary phase in the early Universe.  (Here ``slow-roll'' refers to the regime in which 
the inflaton's kinetic energy remains small compared to its potential energy, yielding an almost constant Hubble parameter.)  Background curvature affects dispersion relations, mode mixing, and asymptotic behavior of gravitational waves, which motivates studying the massive scalar mode and tensor modes directly on de Sitter space~\cite{Higuchi_1987,Bonga2015}.  Because GR with a cosmological constant supports only two tensor polarizations, de Sitter space provides a clean setting for isolating any additional modes arising from $f(R)$ gravity and for making cosmological links between inflationary physics and late-time acceleration.

The structure of this paper is as follows.  In Section~2 we derive the field equations of metric $f(R)$ gravity on a de Sitter background and obtain the Klein--Gordon equation for the extra scalar mode with mass $m_{\psi}$.  Section~3 develops the perturbation of the Ricci tensor, and the resulting linearized field equations. Section~4 performs the (3+1) irreducible decomposition into scalar, vector, and tensor sectors.  In Section~5 we specialize to the model $f(R)=R+\alpha R^{2}-2\Lambda$ and present its explicit linearized dynamics. Section~6 identifies the gauge-invariant Bardeen variables and derives the physical polarization content.  Finally, in Section~7 we verify these results using the geodesic deviation equation, demonstrating that the obtained polarization modes are physically realized in the relative acceleration of 
freely falling test particles.

\section{Field Equations of Metric \texorpdfstring{$f(R)$}{f(R)}
 Gravity on a de Sitter Background}
\label{sec:field-eq-fR}

In metric $f(R)$ gravity, the Einstein--Hilbert Lagrangian density $R$ is replaced by a general function $f(R)$,
\begin{equation}
  S = \frac{1}{2\kappa} \int d^4x \sqrt{-g}\, f(R) + S_m[g_{\mu\nu},\Psi_m],
  \label{eq:action-fR}
\end{equation}
where $S_m$ is the matter action, $\Psi_m$ collectively denotes the matter fields, and
\begin{equation}
  \kappa \equiv 8\pi G = 2M_{\rm Pl}^{-2}
\end{equation}
in terms of the reduced Planck mass $M_{\rm Pl}$ (we use units $c=1$). Varying \eqref{eq:action-fR} with respect to the metric and following, e.g., \cite{sym16081007}, one obtains the metric $f(R)$ field equations
\begin{equation}
  f'(R)\,R_{\mu\nu}
  - \frac{1}{2} g_{\mu\nu} f(R)
  + \bigl(g_{\mu\nu}\Box - \nabla_\mu \nabla_\nu \bigr)f'(R)
  = \kappa\, T_{\mu\nu},
  \label{eq:fR-field-eq}
\end{equation}
where $f'(R) \equiv df/dR$, $\Box \equiv g^{\rho\sigma}\nabla_\rho \nabla_\sigma$, and
\begin{equation}
  T_{\mu\nu}
  = -\frac{2}{\sqrt{-g}} \frac{\delta (\sqrt{-g}\, S_m)}{\delta g^{\mu\nu}}
  \label{eq:Tmunu-def}
\end{equation}
is the matter stress--energy tensor. In vacuum we set $T_{\mu\nu}=0$ and \eqref{eq:fR-field-eq} reduces to
\begin{equation}
  f'(R)\,R_{\mu\nu}
  - \frac{1}{2} g_{\mu\nu} f(R)
  + \bigl(g_{\mu\nu}\Box - \nabla_\mu \nabla_\nu \bigr)f'(R)
  = 0.
  \label{eq:fR-field-eq-vac}
\end{equation}

Taking the trace of \eqref{eq:fR-field-eq} yields the scalar (trace) equation
\begin{equation}
  3\Box f'(R) + R f'(R) - 2 f(R) = \kappa T,
  \label{eq:trace-fR}
\end{equation}
where $T \equiv g^{\mu\nu}T_{\mu\nu}$. In vacuum,
\begin{equation}
  3\Box f'(R) + R f'(R) - 2 f(R) = 0.
  \label{eq:trace-fR-vac}
\end{equation}
The trace equation will be the starting point for identifying the massive scalar propagating mode (the ``scalaron'') and its effective mass.
%-------------------------------------------------------------
\subsection{\texorpdfstring{$f(R)$}{f(R)} gravity, scalar--tensor form, and the chameleon mechanism}
\label{subsec:chameleon}
%-------------------------------------------------------------

The field equations \eqref{eq:fR-field-eq} contain higher derivatives of the metric through
$\Box f'(R)$ and $\nabla_\mu\nabla_\nu f'(R)$. A useful way to expose the extra scalar degree of
freedom and to analyze screening—namely, the suppression of scalar-mediated fifth forces in high-density environments via an environment-dependent effective mass—is to recast metric
$f(R)$ gravity as a scalar--tensor theory via a conformal transformation (see, e.g.,
\cite{DeFelice2010,Sotiriou2010}).

Define
\begin{equation}
  F(R) \equiv f'(R) > 0,
  \qquad
  F(R) = \exp\!\Bigl(-\frac{2\beta \phi}{M_{\rm Pl}}\Bigr),
\end{equation}
where $\phi$ is a scalar field. In metric $f(R)$ gravity the coupling parameter is fixed to
$\beta = 1/\sqrt{6}$, reflecting the universal strength of the scalar coupling to matter. We
then introduce the conformal transformation
\begin{equation}
  \bar{g}_{\mu\nu} = F(R)\, g_{\mu\nu}
  = \exp\!\Bigl(-\frac{2\beta \phi}{M_{\rm Pl}}\Bigr) g_{\mu\nu},
  \label{eq:conf-transform}
\end{equation}
which maps the \emph{Jordan frame} metric $g_{\mu\nu}$ to the \emph{Einstein frame} metric
$\bar{g}_{\mu\nu}$. In the Jordan frame, matter is minimally coupled to $g_{\mu\nu}$ and freely
falling test particles follow geodesics of $g_{\mu\nu}$. In the Einstein frame, the
gravitational sector takes the Einstein--Hilbert form plus a canonical scalar field, while
matter acquires a $\phi$-dependent non-minimal coupling with respect to the Einstein-frame
metric $\bar{g}_{\mu\nu}$.

In terms of $\bar{g}_{\mu\nu}$ and $\phi$, the action \eqref{eq:action-fR} becomes
\begin{equation}
  S = \int d^4x \sqrt{-\bar{g}}
  \biggl[
    \frac{M_{\rm Pl}^2}{2}\,\bar{R}
    - \frac{1}{2}\,\bar{g}^{\mu\nu}
      \bar{\nabla}_\mu \phi\, \bar{\nabla}_nu \phi
    - V(\phi)
  \biggr]
  + S_m\bigl(A^2(\phi)\,\bar{g}_{\mu\nu},\Psi_m\bigr),
  \label{eq:Einstein-frame-action}
\end{equation}
where $\bar{\nabla}_\mu$ and $\bar{R}$ are the covariant derivative and Ricci scalar of
$\bar{g}_{\mu\nu}$, and
\begin{equation}
  A^2(\phi) = F^{-1}(R)
  = \exp\!\Bigl(+\frac{2\beta\phi}{M_{\rm Pl}}\Bigr)
\end{equation}
is the conformal factor relating the Einstein and Jordan metrics in the matter sector. The
scalar potential is
\begin{equation}
  V(\phi)
  = \frac{M_{\rm Pl}^2}{2}
    \frac{R F(R) - f(R)}{F^2(R)}
  = \frac{1}{\kappa}
    \frac{R f'(R) - f(R)}{f'^2(R)}.
  \label{eq:potential-phi}
\end{equation}

The corresponding Einstein-frame field equations can be written as
\begin{align}
  \bar{G}_{\mu\nu}
  &= \frac{1}{M_{\rm Pl}^2}\Bigl(
      T^{(\phi)}_{\mu\nu}
      + T^{(m)}_{\mu\nu}
    \Bigr), \\
  \bar{\Box}\phi
  &= V'(\phi) - \frac{\beta}{M_{\rm Pl}}\,T^{(m)},
  \label{eq:Einstein-frame-eqs}
\end{align}
where $T^{(\phi)}_{\mu\nu}$ is the scalar-field stress tensor and $T^{(m)}_{\mu\nu}$ is the
Einstein-frame matter tensor. For a spatially homogeneous scalar field $\phi=\phi(t)$ in a
spatially flat FRW background, this equation reduces to the standard cosmological
Klein--Gordon equation
\begin{equation}
  \ddot{\phi}+3H\dot{\phi}+V'(\phi)
  = \frac{\beta}{M_{\rm Pl}}\,T^{(m)},
  \label{eq:KG-FRW}
\end{equation}
which in the vacuum limit ($T^{(m)}=0$) becomes
\begin{equation}
  \ddot{\phi}+3H\dot{\phi}+V'(\phi)=0,
  \label{eq:KG-vacuum}
\end{equation}
the equation solved in inflationary and background cosmological applications
\cite{Mukhanov2005,Weinberg2008,DeFelice2010}. This explicitly links the early-time
inflationary dynamics of the model to its late-time cosmological implications.

\paragraph{Jordan vs Einstein frame in practice.}
The Jordan frame is the one in which matter is minimally coupled and experimental observables
(such as test-particle trajectories and detector responses) are most directly interpreted.
The Einstein frame is mathematically convenient for analyzing the dynamics of the extra scalar
degree of freedom and for discussing stability, screening mechanisms, and cosmological
evolution. Physical predictions are frame-independent provided one consistently transforms
both the metric and matter variables. In this paper, we perform the gravitational-wave
analysis in the Jordan frame (where the Bardeen variables and metric perturbations are
defined), while the Einstein-frame description is used only to clarify the scalar--tensor
structure and the chameleon mechanism.

\paragraph{Chameleon mechanism in \texorpdfstring{$f(R)$}{f(R)} gravity.}
The additional scalar degree of freedom in metric \texorpdfstring{$f(R)$}{f(R)} gravity
mediates a universal fifth force \cite{Sotiriou2010, DeFelice2010} through its coupling to
the trace of the matter stress--energy tensor. The scalar field $\phi$ in
\eqref{eq:Einstein-frame-action} couples universally to matter via
$A^2(\phi)\,\bar{g}_{\mu\nu}$ and can mediate this fifth force unless its effective mass
becomes large in high-density environments. The \emph{chameleon mechanism} exploits the
density dependence of the effective potential
\begin{equation}
  V_{\rm eff}(\phi)
  = V(\phi) + \rho\,A(\phi),
  \qquad A(\phi) = \exp\!\Bigl(+\frac{\beta\phi}{M_{\rm Pl}}\Bigr),
  \label{eq:Veff-phi}
\end{equation}
where $\rho$ is the local matter density. In regions of high density, $V_{\rm eff}$ develops a
minimum at which the effective mass
\begin{equation}
  m_{\rm eff}^2(\rho)
  \equiv \left.\frac{d^2V_{\rm eff}}{d\phi^2}\right|_{\phi_{\rm min}(\rho)}
\end{equation}
is large, so that the scalar-mediated force is short-ranged and consistent with local tests
of gravity. In low-density environments (cosmological scales) the minimum shifts and
$m_{\rm eff}$ can become small enough for the scalar to drive cosmic acceleration or leave
imprints on structure formation \cite{Khoury:2003aq,Brax:2008hh,Burrage:2017qrf}.

For an $f(R)$ model to exhibit viable chameleon behavior, the scalar potential $V(\phi)$
derived from \eqref{eq:potential-phi} must satisfy certain conditions in at least part of
field space,
\begin{equation}
  V'(\phi) < 0,
  \qquad
  V''(\phi) > 0,
  \qquad
  V'''(\phi) < 0,
\end{equation}
which translate into nontrivial constraints on the form of $f(R)$ and its derivatives
\cite{Brax:2008hh}. This ensures that the scalar field can be heavy in high-density regions
while remaining light enough on cosmological scales to influence late-time acceleration.

%-------------------------------------------------------------
\subsection{Slow-roll inflation and scalaron dynamics in the Einstein frame}
\label{subsec:slow-roll}
%-------------------------------------------------------------

The scalar--tensor reformulation of metric $f(R)$ gravity introduced above provides a natural
framework for discussing early-Universe inflation driven by the scalaron. When written in the Einstein frame, the scalar field $\phi$ obeys the field equation obtained by varying the
Einstein-frame action \eqref{eq:Einstein-frame-action},
\begin{equation}
    \bar{\Box}\phi = V'(\phi),
\end{equation}
where $V(\phi)$ is given in Eq.~\eqref{eq:potential-phi}. For a spatially homogeneous scalar
field $\phi=\phi(t)$ evolving in a spatially flat FLRW background,
$d\bar{s}^2=-dt^2+a^2(t)d\vec{x}^{\,2}$, this equation reduces to the standard cosmological
Klein--Gordon equation \cite{Mukhanov2005,Weinberg2008,DeFelice2010}
\begin{equation}
    \ddot{\phi} + 3H\dot{\phi} + V'(\phi) = 0,
    \label{eq:KG-Einstein-frame}
\end{equation}
where $H=\dot{a}/a$. The background expansion is governed by the Friedmann equation
\begin{equation}
    3 M_{\rm Pl}^2 H^2 = \frac{1}{2}\dot{\phi}^2 + V(\phi).
\end{equation}

Inflation occurs when the potential dominates the kinetic energy and the field slowly rolls
along $V(\phi)$. This is quantified by the slow-roll parameters
\begin{equation}
    \epsilon(\phi) \equiv \frac{M_{\rm Pl}^2}{2}
        \left( \frac{V'(\phi)}{V(\phi)} \right)^2,
    \qquad
    \eta(\phi) \equiv
        M_{\rm Pl}^2 \frac{V''(\phi)}{V(\phi)},
    \label{eq:slow-roll-parameters}
\end{equation}
where inflation requires $\epsilon \ll 1$ and $|\eta| \ll 1$. Under these conditions,
Eq.~\eqref{eq:KG-Einstein-frame} reduces to the familiar slow-roll equation
\begin{equation}
    3H\dot{\phi} \simeq -V'(\phi),
    \label{eq:slow-roll-eom}
\end{equation}
and the Hubble parameter satisfies $H^2 \simeq V(\phi)/(3M_{\rm Pl}^2)$.

For the Starobinsky-type models considered in this work, the potential $V(\phi)$ possesses a
nearly flat region at large curvature ($R \gg \Lambda$), ensuring that the slow-roll
conditions \eqref{eq:slow-roll-parameters} are naturally satisfied. In this regime the scalaron behaves as an inflaton with an effective mass $m_\phi^2 \simeq V''(\phi)$, and the inflationary predictions coincide with those of the well-known $R+\alpha R^2$ model
\cite{Starobinsky1980,Starobinsky1983}. Observationally, the slow-roll phase gives rise to a nearly scale-invariant spectrum of primordial curvature perturbations and a suppressed tensor-to-scalar ratio, in excellent agreement with current CMB constraints.

Although slow-roll inflation operates at curvature scales far above those relevant for present-day gravitational-wave detectors, the same underlying extra scalar degree of freedom of $f(R)$ gravity governs both regimes. In the inflationary context this degree of freedom is commonly referred to as the \emph{scalaron}, while at the level of linear perturbations it appears as a propagating massive scalar mode. In particular, the mass of the scalar perturbation at the de Sitter solution,
\begin{equation}
m_\psi^2 = \frac{1}{3}\left[ \frac{f'(R_d)}{f''(R_d)} - R_d \right],
\end{equation}
controls the propagation of the scalar polarization of gravitational waves and provides a link between the early-time inflationary dynamics of the model and its late-time cosmological implications.

\paragraph{de Sitter solutions and cosmological implications.}
We are particularly interested in constant-curvature de Sitter solutions and small perturbations around them. For a vacuum constant-curvature background with $R = R_d = \mathrm{const}$ and $T=0$, the trace equation \eqref{eq:trace-fR-vac} reduces to the algebraic condition
\begin{equation}
  R_d f'(R_d) - 2 f(R_d) = 0.
  \label{eq:deS-condition}
\end{equation}
Any function $f(R)$ that admits a solution of \eqref{eq:deS-condition} possesses a de Sitter solution with curvature $R_d$. (Anti--de Sitter solutions correspond to constant-curvature
solutions with $R_d<0$ and must be analyzed separately.) For the modified Starobinsky model
\begin{equation}
  f(R) = R + \alpha R^2 - 2\Lambda,
  \label{eq:Starobinsky-modified}
\end{equation}
the de Sitter curvature $R_d$ is determined by
\begin{equation}
  R_d (1 + 2\alpha R_d) - 2(R_d + \alpha R_d^2 - 2\Lambda) = 0,
\end{equation}
which reduces to $R_d \simeq 4\Lambda$ in the late-time, low-curvature regime
$\alpha R_d \ll 1$.\footnote{At very high curvature (early Universe) the $\alpha R^2$ term
dominates and the model approaches the inflationary Starobinsky regime, whereas at low curvature (late times) the cosmological constant term $-2\Lambda$ drives accelerated expansion.} Thus, constant-curvature solutions in $f(R)$ gravity provide a unified framework for modeling both early-time inflation and late-time dark-energy--dominated epochs, and they form the natural background for our gravitational-wave polarization analysis.

In the remainder of this paper, we will work in the Jordan frame and treat the extra propagating scalar mode directly in terms of the curvature perturbation $\delta R$. To avoid confusion of notation, we will use:
\begin{itemize}
  \item $\phi$ for the Einstein-frame scalar field entering the scalar--tensor and inflationary description in this subsection; and
  \item $\delta R$ (or, equivalently, a canonically normalized scalar perturbation $\psi$ with
        mass $m_\psi$) for the extra propagating scalar mode that appears in the linearized Jordan-frame field equations and in the Bardeen-variable analysis.
\end{itemize}

%-------------------------------------------------------------
\subsection{Metric perturbations around a de Sitter background}
\label{subsec:metric-perturbations}
%-------------------------------------------------------------

We now consider small perturbations around a background solution $g_{\mu\nu}$ which solves the vacuum field equations \eqref{eq:fR-field-eq-vac}. In particular, we will later specialize to a de Sitter background satisfying \eqref{eq:deS-condition}. The perturbed metric is written as
\begin{equation}
  g_{\mu\nu} \;\longrightarrow\;
  \tilde{g}_{\mu\nu} = g_{\mu\nu} + h_{\mu\nu},
  \qquad
  \tilde{g}^{\mu\nu} = g^{\mu\nu} - h^{\mu\nu} + \mathcal{O}(h^2),
  \label{eq:metric-perturbation}
\end{equation}
where indices on $h_{\mu\nu}$ are raised and lowered with the background metric $g_{\mu\nu}$.

To linear order, the curvature quantities and the function $f(R)$ expand as
\begin{align}\label{eq:fR-expansions}
  \tilde{R}
  &= R + \delta R + \mathcal{O}(h^2), \nonumber\\
  \tilde{f}(R)
  &= f(R) + f'(R)\,\delta R + \mathcal{O}(h^2), \\
  \tilde{f}'(R)
  &= f'(R) + f''(R)\,\delta R + \mathcal{O}(h^2), \nonumber
\end{align}
where $R$ is the background Ricci scalar and $\delta R$ is its perturbation.

It is convenient to separate background and perturbed covariant derivatives. Denoting by
$\tilde{\nabla}_\mu$ the covariant derivative associated with $\tilde{g}_{\mu\nu}$ and by
$\nabla_\mu$ the one associated with $g_{\mu\nu}$, the difference between them acting on a
generic tensor $T^{\nu_1\cdots\nu_k}{}_{\rho_1\cdots\rho_l}$ is (see, e.g.,
\cite{Wald:1984rg})
\begin{equation}
  \tilde{\nabla}_\mu T^{\nu_1\cdots\nu_k}{}_{\rho_1\cdots\rho_l}
  = \nabla_\mu T^{\nu_1\cdots\nu_k}{}_{\rho_1\cdots\rho_l}
  + \sum_i C^{\nu_i}{}_{\mu\sigma}\,
      T^{\nu_1\cdots\sigma\cdots\nu_k}{}_{\rho_1\cdots\rho_l}
  - \sum_j C^{\sigma}{}_{\mu\rho_j}\,
      T^{\nu_1\cdots\nu_k}{}_{\rho_1\cdots\sigma\cdots\rho_l},
  \label{eq:covariant-derivative-variation}
\end{equation}
where the connection difference $C^\rho{}_{\mu\nu}$ is
\begin{equation}
  C^\rho{}_{\mu\nu}
  = \frac{1}{2}\,\tilde{g}^{\rho\sigma}
    \bigl(
      \nabla_\mu \tilde{g}_{\nu\sigma}
      + \nabla_\nu \tilde{g}_{\mu\sigma}
      - \nabla_\sigma \tilde{g}_{\mu\nu}
    \bigr).
  \label{eq:C-connection}
\end{equation}
All quantities without tildes refer to the background metric. These relations allow one to
express perturbed curvature tensors and the perturbed trace equation in terms of $h_{\mu\nu}$
and $\delta R$.

Varying the vacuum trace equation \eqref{eq:trace-fR-vac} and keeping terms linear in the
perturbations yields
\begin{align}
  0
  &= \delta\!\left[
      \Box f'(R) + \frac{1}{3}R f'(R) - \frac{2}{3} f(R)
    \right] \nonumber\\
  &= \Biggl[
       \Box f''(R)
       + f''(R)\Box
       - \frac{1}{3}
         \bigl(
           f'(R) - R f''(R)
         \bigr)
     \Biggr]\delta R \nonumber\\
  &\quad
     - \Biggl[
       h^{\mu\rho}\nabla_\mu
       + \frac{1}{2} g^{\mu\nu} g^{\rho\sigma}
         \bigl(
           \nabla_\mu h_{\nu\sigma}
           + \nabla_\nu h_{\mu\sigma}
           - \nabla_\sigma h_{\mu\nu}
         \bigr)
     \Biggr]
     \bigl(f''(R)\,\partial_\rho R\bigr),
  \label{eq:delta-trace-general}
\end{align}
where we have used \eqref{eq:fR-expansions} and
\eqref{eq:covariant-derivative-variation}. Equation \eqref{eq:delta-trace-general} is valid
for a general background.

For the de Sitter backgrounds of interest in this work, the Ricci scalar is constant,
\begin{equation}
  R = R_d = \mathrm{const},
\end{equation}
so that $\nabla_\mu R = 0$ and $f'(R_d)$, $f''(R_d)$ are constants. In this case the second line
of \eqref{eq:delta-trace-general} vanishes, $\Box f''(R_d) = 0$, and we obtain the simplified
scalar perturbation equation
\begin{equation}
  \Biggl[
    \Box
    - \frac{1}{3}
      \Bigl(
        \frac{f'(R_d)}{f''(R_d)} - R_d
      \Bigr)
  \Biggr]\delta R = 0,
  \label{eq:deltaR-KG-form}
\end{equation}
or, equivalently,
\begin{equation}
  \bigl(\Box - m_\psi^2\bigr)\,\delta R = 0,
  \label{eq:deltaR-KG}
\end{equation}
with
\begin{equation}
  m_\psi^2
  = \frac{1}{3}\biggl[
      \frac{f'(R_d)}{f''(R_d)} - R_d
    \biggr].
  \label{eq:mpsi-def}
\end{equation}
Here $m_\psi$ is the effective mass of the scalar propagating mode
associated with the curvature perturbation $\delta R$ in the de Sitter background. Equation \eqref{eq:deltaR-KG} is a Klein--Gordon equation for $\delta R$ and describes the propagation of a massive scalar mode in addition to the usual tensor modes of general relativity. In later sections we will relate $\delta R$ to a gauge-invariant Bardeen combination and denote the corresponding massive scalar propagating mode by $\psi$.

\paragraph{Cosmological interpretation of \texorpdfstring{$m_\psi$}{mψ}.}
The mass scale $m_\psi^{-1}$ determines the range of the scalar-mediated interaction and the characteristic dispersion of the scalar polarization of gravitational waves in a de Sitter background. On sub-horizon scales with $k \gg a m_\psi$, the scalar mode behaves effectively massless and can, in principle, contribute to additional polarization signatures. On scales $k \ll a m_\psi$ the mode is strongly suppressed, consistent with local gravity constraints.
For the modified Starobinsky model \eqref{eq:Starobinsky-modified}, one finds $m_\psi^2 \approx 1/(6\alpha)$ in the high-curvature regime, linking the mass scale to both early-time inflationary dynamics and late-time modifications of gravitational-wave propagation in cosmology.

The detailed decomposition of the metric perturbations into scalar, vector, and tensor Bardeen variables, and the identification of the corresponding polarization modes, will be carried out in the following sections.
%-------------------------------------------------------------
\section{Perturbations of the Ricci Tensor \texorpdfstring{$\delta R_{\mu\nu}$}{deltaRmunu} and Scalar Dynamics in \texorpdfstring{$f(R)$}{fR} Gravity}
\label{sec:perturbed-ricci}
%-------------------------------------------------------------

The evolution of cosmological perturbations in $f(R)$ gravity influences both the expansion history of the Universe and the propagation of gravitational waves across different cosmological epochs \cite{Chaichian}. To understand how the gravitational field responds to small deviations from a background metric—whether a cosmological FRW background, a black hole spacetime, or, as in this work, a de Sitter background—it is necessary to compute the perturbation of curvature quantities. Since the Ricci tensor enters directly in the field equations, its perturbation represents the leading-order correction to the spacetime
curvature and is essential for identifying the massive scalar propagating mode present in $f(R)$ theories.

Furthermore, gauge transformations in $f(R)$ gravity are complicated by the presence of higher derivatives of $R$. A fully gauge-invariant description of perturbations therefore requires determining how the scalar curvature perturbation $\delta R$ interacts with metric perturbations $h_{\mu\nu}$ through $\delta R_{\mu\nu}$. This provides a crucial intermediate step on the way to constructing the gauge-invariant Bardeen potentials in later sections.

To obtain the perturbed field equation, we expand the metric as
$g_{\mu\nu}=g^d_{\mu\nu}+h_{\mu\nu}$ and linearize each term of the
$f(R)$ field equation \eqref{eq:fR-field-eq-vac}. Using
$\delta f=f'(R_d)\delta R$ and $\delta f'=f''(R_d)\delta R$, and recalling that $R_d$ is constant, the variation of covariant derivative terms such as
$\nabla_\mu\nabla_\nu f'(R)$ must be treated carefully. In general,
\begin{equation}
\delta(\nabla_\mu\nabla_\nu f')
= \nabla_\mu\nabla_\nu \delta f'
- (\delta \Gamma^\lambda_{\mu\nu})\,\nabla_\lambda f' .
\end{equation}
However, on a constant-curvature de Sitter background one has
\begin{equation}
\nabla_\lambda f'(R_d)=0 ,
\end{equation}
so the connection-variation term vanishes identically. As a result, at linear order
\begin{equation}
\delta(\nabla_\mu\nabla_\nu f')
= \nabla_\mu\nabla_\nu \delta f' .
\end{equation}
After accounting for this simplification, the variation of
$(g_{\mu\nu}\Box - \nabla_\mu\nabla_\nu)f'$ yields the operator
$(g^d_{\mu\nu}\Box - \nabla_\mu\nabla_\nu)\delta f'$.

Perturbing the vacuum field equation \eqref{eq:fR-field-eq-vac} using a de Sitter background ($R = R_d = \mathrm{const}$) yields the linearized equation
\begin{equation}
    f'(R_d)\,\delta R_{\mu\nu}
    + R^{d}_{\mu\nu}\,\delta f'(R_d)
    - \frac{1}{2}\bigl[g^{d}_{\mu\nu}\,\delta f(R_d)
        + h_{\mu\nu}\, f(R_d)\bigr]
    + \bigl(g^{d}_{\mu\nu}\Box - \nabla_\mu\nabla_\nu\bigr)\delta f'(R_d)
    = 0,
    \label{eqn26}
\end{equation}
where $g^{d}_{\mu\nu}$ is the background de Sitter metric.

To obtain \eqref{eqn26}, we decompose the metric as
\begin{equation}
    g_{\mu\nu} = g^{d}_{\mu\nu} + h_{\mu\nu},
    \label{eqn27}
\end{equation}
with indices on $h_{\mu\nu}$ raised and lowered using $g^{d}_{\mu\nu}$. Since de Sitter space is a constant-curvature solution of metric $f(R)$ gravity, the background satisfies
\begin{equation}
    \nabla_\mu f'(R_d) = 0,
    \label{eqn29}
\end{equation}
which follows directly from the constancy of $R_d$. This condition reflects the fact that $f'(R_d)$ is a nonzero constant fixed by the chosen $f(R)$ model and represents the effective gravitational coupling on the background.

The nonperturbed trace equation \eqref{eq:trace-fR-vac} evaluated on a constant-curvature vacuum background yields the algebraic de Sitter condition
\begin{equation}
    R_d f'(R_d) = 2 f(R_d),
    \label{eqn30}
\end{equation}
which determines the allowed background curvature $R_d$ of the $f(R)$ theory. Substituting \eqref{eqn30} into the background field equation \eqref{eq:fR-field-eq-vac} and using $R = R_d$ and $T_{\mu\nu}=0$ gives
\begin{equation}
    R^{d}_{\mu\nu} f'(R_d)
    - \frac{1}{2} g^{d}_{\mu\nu} f(R_d)
    = 0,
    \label{eqn31}
\end{equation}
which implies
\begin{equation}
    R^{d}_{\mu\nu}
    = \frac{1}{2}\frac{f(R_d)}{f'(R_d)}\, g^{d}_{\mu\nu}.
    \label{eqn32}
\end{equation}
Using \eqref{eqn30} once more yields
\begin{equation}
    R^{d}_{\mu\nu} = \frac{1}{4} R_d\, g^{d}_{\mu\nu},
    \label{eqn33}
\end{equation}
showing that the background is an Einstein space and, in fact, maximally symmetric.

For a spatially flat FLRW spacetime, the Ricci scalar is
$R = 6(2H^{2} + \dot{H})$. In de Sitter space $\dot{H}=0$, so
$R_d = 12H_d^{2}$ and $H_d = \sqrt{R_d/12}$.

A de Sitter Universe undergoes exponential expansion,
\begin{equation}
    a(t) \propto e^{H_d t},
    \qquad
    H_d = \sqrt{R_d/12},
\end{equation}
and this constant-curvature solution will serve as the background for our perturbative analysis.

Dividing \eqref{eqn26} by $f'(R_d)$ and substituting
$\delta f = f'_d \delta R$ and $\delta f' = f''_d \delta R$ gives
\begin{align}
    \delta R_{\mu\nu}
    + \Bigl(\frac{f''_d}{f'_d} R^{d}_{\mu\nu}
       - \frac{1}{2} g^{d}_{\mu\nu}\Bigr)\delta R
    - \frac{1}{2}\frac{f_d}{f'_d} h_{\mu\nu}
    + \frac{f''_d}{f'_d}
      \bigl(g^{d}_{\mu\nu}\Box - \nabla_\mu\nabla_\nu\bigr)\delta R
    = 0.
    \label{linearized-field-eqn}
\end{align}

At this stage, it is important to clarify the fate of the algebraic term
$-\frac{1}{2}\frac{f_d}{f'_d} h_{\mu\nu}$ appearing in
Eq.~\eqref{linearized-field-eqn}. On a constant--curvature de~Sitter background, the trace condition
\eqref{eqn30} implies
\begin{equation}
    \frac{f_d}{f'_d} = \frac{R_d}{2},
\end{equation}
so that this contribution may be written as $-(R_d/4)\,h_{\mu\nu}$.
This term is proportional to the background curvature scale and contains no derivatives. For gravitational waves of wavelength $\lambda \ll H_d^{-1}$, corresponding to the local inertial (short--wavelength) limit relevant for detector-scale propagation, such curvature-suppressed algebraic terms do not contribute to
the dynamical wave equation. They may therefore be consistently neglected, or equivalently absorbed into the background de~Sitter curvature. With this understanding, the linearized field equation reduces to Eq.~\eqref{eqn38}.

To evaluate \eqref{linearized-field-eqn}, we recall the general linearized Ricci tensor
\cite{Carroll:2004st,Wald:1984rg}
\begin{equation}
    \delta R_{\mu\nu}
    = \frac{1}{2}
        \bigl(
            -\Box h_{\mu\nu}
            - \nabla_\mu\nabla_\nu h
            + \nabla^\rho\nabla_\mu h_{\rho\nu}
            + \nabla^\rho\nabla_\nu h_{\rho\mu}
        \bigr)
    + \mathcal{O}(h^2),
    \label{eq:deltaRmunu}
\end{equation}
where $h \equiv h^\rho{}_\rho$.

For gravitational-wave propagation at detector scales, the wavelength of the perturbation is much shorter than the de Sitter curvature radius $H_d^{-1}$. One may therefore work in the
local inertial frame of the background, in which
\begin{equation}
    g_{\mu\nu} \approx \eta_{\mu\nu}, \qquad R_{\mu\nu} \approx 0,
    \label{eqn36}
\end{equation}
while still retaining the nonzero constants $f'_d$ and $f''_d$. In this limit, curvature-suppressed algebraic terms proportional to $h_{\mu\nu}$ do not contribute to the dynamical propagation of high-frequency gravitational waves.

Under this approximation, \eqref{linearized-field-eqn} reduces to
\begin{equation}
    \delta R_{\mu\nu}
    - \frac{1}{2}\eta_{\mu\nu}\,\delta R
    + \frac{f''_d}{f'_d}
      (\eta_{\mu\nu}\Box - \partial_\mu\partial_\nu)\delta R
    = 0.
    \label{eqn38}
\end{equation}

The linearized scalar curvature is
\begin{equation}
    \delta R \equiv R[h],
    \label{eqn39}
\end{equation}
and the linearized Einstein tensor reduces to
\begin{equation}
    \delta G_{\mu\nu}
    = \delta R_{\mu\nu}
      - \frac{1}{2}\eta_{\mu\nu}\,\delta R[h].
    \label{eqn40}
\end{equation}
Substituting \eqref{eqn39} and \eqref{eqn40} into \eqref{eqn38} yields the perturbed field
equation
\begin{equation}
    \delta G_{\mu\nu}
    + \frac{1}{3 m_\psi^{2}}
      (\eta_{\mu\nu}\Box - \partial_\mu\partial_\nu)\,\delta R
    = 0,
    \label{perturbed-field-eqn-vacuum}
\end{equation}
where the effective scalar mass $m_\psi$ arises from the trace of the linearized field equations and is given by
\begin{equation}
    m_\psi^2
    = \frac{1}{3}
      \Bigl(
          \frac{f'_d}{f''_d} - R_d
      \Bigr).
    \label{eqn42}
\end{equation}
In the Minkowski limit $(R_d \rightarrow 0)$, this reduces to
\begin{equation}
    m_\psi^2 = \frac{1}{3}\frac{f'_d}{f''_d}.
    \label{ms-equation}
\end{equation}

Once the effective scalar mass $m_\psi^{2}$ is identified, its physical meaning becomes transparent by considering a Fourier (plane-wave) decomposition of the gauge-invariant variables. In a constant-curvature background, linear perturbations admit the ansatz
$X(t,\mathbf{x}) = X_k e^{-i(\omega t - \mathbf{k}\cdot\mathbf{x})}$, which diagonalizes the spatial Laplacian. Substituting this into the Klein--Gordon--type equation
$(\Box - m_\psi^{2})\,\delta R = 0$ yields the dispersion relation
$\omega^{2} = k^{2} + m_\psi^{2}$. Thus the extra scalar mode propagates as a massive mode, in contrast to the transverse tensor polarizations, which remain massless. This plane-wave form clarifies how the additional scalar polarization arises in metric $f(R)$ gravity.

These expressions show that the scalar curvature perturbation $\delta R$ propagates as a massive scalar field on the de Sitter background. The mass $m_\psi$ controls the range and dispersion of the scalar gravitational-wave mode and depends explicitly on the background curvature $R_d$. Since $R_d$ changes across cosmological epochs, the behavior of the scalar mode encodes information about the underlying cosmic expansion and offers potential observational signatures beyond the standard tensor modes.

%-------------------------------------------------------------
\section{3+1 Decomposition and the Scalar, Vector, Tensor Modes of \texorpdfstring{$f(R)$}{f(R)} Gravity}
\label{sec:3p1}
%-------------------------------------------------------------

In this section we analyze the scalar, vector, and tensor perturbations of the metric using the
standard scalar--vector--tensor (SVT) decomposition. As discussed in
Section~\ref{sec:field-eq-fR}, gravitational waves detected at astrophysical scales propagate on a
spacetime whose curvature radius is much larger than their wavelength.
Therefore, for the purpose of the 3+1 decomposition we work in the local
Minkowski limit of the de~Sitter background,
\begin{equation}
    g_{\mu\nu} \approx \eta_{\mu\nu},
\end{equation}
while retaining the constant background quantities $f'_d$, $f''_d$, and the mass $m_\psi$ of massive scalar propagating mode. Throughout this section we adopt the metric signature $(+---)$. For the scalar sector we work in the longitudinal (Newtonian) gauge, in which the metric perturbations are encoded in the gauge-invariant Bardeen potentials $\Phi$ and $\Psi$.

%-------------------------------------------------------------
\subsection{Scalar mode}
%-------------------------------------------------------------

The $00$ component of the linearized Einstein tensor in the longitudinal gauge is
\begin{equation}\label{eqn44}
    \delta G_{00} = -2\nabla^2\Phi.
\end{equation}
The $00$ component of the perturbed $f(R)$ field equation
\eqref{perturbed-field-eqn-vacuum} then takes the form
\begin{equation}\label{eqn45}
    \delta G_{00}
    - \frac{1}{3m_\psi^2}\,\nabla^{2}\delta R
    = 0.
\end{equation}
Substituting \eqref{eqn44} into \eqref{eqn45} yields
\begin{align}
    -2\nabla^2\Phi
    - \frac{1}{3m_\psi^2}\nabla^{2}\delta R = 0,
\end{align}
so that
\begin{equation}
    2\Phi + \frac{\delta R}{3m_\psi^2} = 0,
\end{equation}
and therefore
\begin{equation}\label{48}
    \Phi = -\frac{\delta R}{6 m_\psi^2}.
\end{equation}

The traceless spatial components ($i\neq j$) yield the anisotropy constraint
\begin{equation}
    \Phi - \Psi = \Pi,
\end{equation}
where $\Pi$ denotes the (gauge-invariant) anisotropic stress. In vacuum,
$\Pi=0$, and the two Bardeen potentials coincide,
\begin{equation}\label{49}
    \Phi = \Psi = -\frac{\delta R}{6 m_\psi^2}.
\end{equation}
Thus the curvature perturbation $\delta R$ directly sources the scalar
Bardeen potentials even in the absence of matter anisotropic stress,
producing the massive scalar (breathing/longitudinal) gravitational-wave
polarization predicted in $f(R)$ gravity. In a cosmological context, this
relation links the extra gravitational-wave polarization to the scalar sector
of cosmological perturbations and to the scalar mass $m_\psi$ on a de~Sitter
background.

\paragraph{}
It is worth noting that the expression
$\Phi=\Psi=-\delta R/(6m_\psi^2)$ and the mass parameter $m_\psi^2$ used in
this local 3+1 analysis correspond to the short-wavelength, locally
Minkowskian limit of a de~Sitter background. In this regime the curvature
radius $H^{-1}$ is much larger than the gravitational-wave wavelength, and
terms proportional to the background curvature $R_d$ are negligible.
Consequently, the scalar mass reduces to
$m_\psi^2 \simeq \tfrac{1}{3}f'_d/f''_d = 1/(6\alpha)$. In a fully global
de~Sitter treatment, however, the effective mass contains an additional
curvature contribution and takes the form
$m_\psi^2 = \tfrac{1}{3}(f'(R_d)/f''(R_d) - R_d)$. Thus the local vacuum 3+1
decomposition used here captures the correct propagation physics for
gravitational waves measured in a local inertial frame, while the global
de~Sitter mass governs the long-wavelength, cosmological evolution of the
scalar mode.

%-------------------------------------------------------------
\subsection{Vector modes}
%-------------------------------------------------------------

Vector perturbations appear in the $0i$ components of the metric as
divergence-free vectors. The gauge-invariant combination is
\cite{MukhanovFeldmanBrandenberger1992,KodamaSasaki1984}
\begin{equation}
    V_i \equiv S_i - F_i', \qquad \partial^i V_i = 0,
\end{equation}
where $S_i$ enters the $g_{0i}$ component and $F_i$ the vector part of the
spatial metric.

In the vector sector all scalar perturbations vanish, so in particular
\begin{equation}
    \delta R = 0.
\end{equation}
The linearized Ricci tensor reduces to \cite{Alves_2024,Flanagan_2005}
\begin{equation}
    \delta R_{0i} = -\frac{1}{2}\nabla^2 V_i.
\end{equation}
The perturbed field equations in vacuum imply
\begin{equation}
    \delta G_{0i} = \delta R_{0i} = 0,
\end{equation}
and hence
\begin{equation}
    \nabla^2 V_i = 0.
\end{equation}
Under localized boundary conditions this yields $V_i = 0$. As in general
relativity, no vector modes propagate in vacuum metric $f(R)$ gravity.

%-------------------------------------------------------------
\subsection{Tensor modes}
%-------------------------------------------------------------

We now examine the transverse–traceless (TT) tensor perturbations $h_{ij}^{TT}$. The spatial components of the perturbed $f(R)$ field equations take the form
\begin{equation}\label{eqn54}
    \delta G_{ij}
    + \frac{1}{3 m_\psi^{2}}
      (\delta_{ij}\Box - \partial_i\partial_j)\,\delta R
    = 0.
\end{equation}
The SVT decomposition of the perturbed Ricci tensor is \cite{Flanagan_2005,Alves_2024,MukhanovFeldmanBrandenberger1992,Capozziello_2011}
\begin{equation}\label{eqn55}
    \delta R_{ij}
    = -\partial_i\partial_j(\Phi + \Psi)
      - \delta_{ij}(-\ddot{\Phi} + \nabla^2\Phi)
      - \frac{1}{2}(\partial_i\dot{V}_j + \partial_j\dot{V}_i)
      - \frac{1}{2}\Box h_{ij}^{TT}.
\end{equation}

In the pure tensor sector,
\[
\Phi = \Psi = 0, \qquad V_i = 0,
\qquad \delta R = 0.
\]
Equation \eqref{eqn55} reduces to
\begin{equation}
    \delta R_{ij} = -\frac{1}{2}\Box h_{ij}^{TT}.
\end{equation}
The term $(\delta_{ij}\Box - \partial_i\partial_j)\delta R$ in \eqref{eqn54} has no TT projection and therefore drops out. Substituting into \eqref{eqn54} yields
\begin{equation}
    \Box h_{ij}^{TT} = 0.
\end{equation}
Thus, the tensor modes in metric $f(R)$ gravity propagate exactly as in GR: they satisfy the standard wave equation, travel at the speed of light, and possess only two transverse-traceless polarization states. All deviations from GR in gravitational-wave propagation therefore originate exclusively from the massive scalar propagating mode.

\section{Analyzing a specific \texorpdfstring{$f(R)$}{f(R)} model}
\label{sec: specific f(R)}

We now specialize the general discussion of
Secs.~\ref{sec:field-eq-fR} and \ref{sec:perturbed-ricci} to a concrete and widely studied model. One of the simplest and most successful choices is the Starobinsky model
\begin{equation}\label{eqn58}
    f(R)=R+\alpha R^2+O(R^3),
\end{equation}
which provides a purely geometric mechanism for early-Universe
inflation and introduces an additional scalar degree of freedom through the higher-curvature term \cite{Starobinsky1980,Starobinsky1983}. In this framework inflation is driven by the $R^2$ correction itself, rather than by an independent inflationary field, with the parameter
$\alpha$ setting the characteristic inflationary scale and controlling the amplitude of primordial fluctuations
\cite{MukhanovChibisov1981,Iacconi_2023}.

For the truncated model \eqref{eqn58}, the first and second derivatives of $f(R)$ are
\begin{equation}\label{eqn59}
    f'(R) \approx 1 + 2\alpha R,
\end{equation}
\begin{equation}\label{eqn60}
    f''(R) \approx 2\alpha.
\end{equation}
The vacuum field equation \eqref{eq:fR-field-eq-vac} takes the form
\begin{equation}
    f'(R) R_{\mu\nu}
    -\frac{1}{2}g_{\mu\nu}f(R)
    +\bigl(g_{\mu\nu}\Box-\nabla_{\mu}\nabla_{\nu}\bigr)f'(R)=0,
\end{equation}
with the corresponding trace equation
\begin{equation}\label{Trace eqn}
    3\Box f'(R)+R f'(R)-2f(R)=0.
\end{equation}

%-------------------------------------------------------------
\subsection{Constant-curvature backgrounds and the need for a modified Starobinsky model}
%-------------------------------------------------------------

We now seek constant-curvature vacuum solutions characterized by
\begin{equation}
    g_{\mu\nu} = g_{\mu\nu}^d,
    \qquad
    R = R_d = \text{const},
    \qquad
    \nabla_{\mu}R_d=0.
\end{equation}
For such backgrounds one has
\begin{equation}
    \Box f'(R_d) = 0,
\end{equation}
and the trace equation \eqref{Trace eqn} reduces to the algebraic
condition
\begin{equation}\label{R-d-eqn}
    R_d f'(R_d)-2f(R_d)=0.
\end{equation}
Substituting the Starobinsky form \eqref{eqn58} into
\eqref{R-d-eqn} yields the unique solution
\begin{equation}
    R_d=0.
\end{equation}
Thus, within the pure Starobinsky model \eqref{eqn58}, Minkowski
spacetime is the only constant-curvature vacuum solution. In particular,
there is no nontrivial de~Sitter background with $R_d>0$ that could
describe an exponentially expanding late-time Universe.

From a cosmological standpoint this limitation motivates extending the
model to include a vacuum-energy contribution. We therefore adopt the
\emph{modified} Starobinsky model
\begin{equation}
    f(R)=R+\alpha R^2-2\Lambda,
\end{equation}
which supplements the inflationary $R+\alpha R^2$ sector with a
cosmological constant term. In this model the quadratic curvature term
governs early-time inflation \cite{Starobinsky1980}, while the
constant contribution $-2\Lambda$ drives late-time accelerated
expansion \cite{Sotiriou2010,Weinberg_1989}.

For the modified model, the constant-curvature condition
\eqref{R-d-eqn} yields
\begin{equation}
    R_d=4\Lambda.
\end{equation}
This solution defines a de~Sitter background on which we will linearize
the field equations. As shown in
Sec.~\ref{sec:perturbed-ricci}, a constant-$R_d$ background of this type
satisfies
\begin{equation}
    R_{\mu\nu}\big|_{R_d}
    = \frac{1}{4}R_d\,g_{\mu\nu},
\end{equation}
so the spacetime is an \emph{Einstein space}, characterized by
$R_{\mu\nu}\propto g_{\mu\nu}$. This geometric notion should be
distinguished from the \emph{Einstein frame} discussed in
Sec.~\ref{subsec:chameleon}, which is obtained from the Jordan frame by a
conformal transformation.

From a cosmological perspective, the modified Starobinsky model thus
provides a unified description of the background dynamics: at high
curvature the $\alpha R^2$ term drives inflation, while at low curvature
the $-2\Lambda$ term yields a late-time de~Sitter phase with curvature
$R_d=4\Lambda$. This de~Sitter solution serves as the background for the
perturbative and gravitational-wave analyses developed in the following
sections.
%-------------------------------------------------------------
\subsection{Einstein-frame potential and stability around the de Sitter point}
%-------------------------------------------------------------

In Sec.~\ref{subsec:chameleon} we reviewed the scalar--tensor (Einstein-frame)
representation of metric $f(R)$ gravity. For the modified Starobinsky model
\begin{equation}
    f(R)=R+\alpha R^2-2\Lambda ,
\end{equation}
the Einstein-frame scalar potential is
\begin{equation}
    V(\phi)
    = \frac{M_{\rm Pl}^2}{2}\,
      \frac{\alpha R^2 + 2\Lambda}{(1+2\alpha R)^2},
    \label{eq:Einstein_potential}
\end{equation}
where the scalar field $\phi$ is related to the curvature through the conformal
relation
\begin{equation}
    f'(R)=1+2\alpha R
    = \exp\!\left(-\frac{2\beta\phi}{M_{\rm Pl}}\right),
    \label{eq:conformal_relation}
\end{equation}
with $\beta$ defined as in Sec.~\ref{subsec:chameleon}.  Equation
\eqref{eq:conformal_relation} implicitly defines $R=R(\phi)$.

\paragraph*{Chain rule and curvature derivatives.}
Using \eqref{eq:conformal_relation}, the derivative of $R$ with respect to $\phi$
is
\begin{equation}
    \frac{dR}{d\phi}
    = -\frac{2\beta}{M_{\rm Pl}}\,
      \frac{f'(R)}{f''(R)}
    = -\frac{\beta}{\alpha M_{\rm Pl}}\,(1+2\alpha R),
    \label{eq:dR_dphi}
\end{equation}
where we used $f''(R)=2\alpha$.  All derivatives of $V(\phi)$ follow from repeated
application of the chain rule
\(
d/d\phi = (dR/d\phi)\, d/dR .
\)

\paragraph*{First derivative.}
Differentiating \eqref{eq:Einstein_potential} with respect to $R$ gives
\begin{equation}
    \frac{dV}{dR}
    = \frac{M_{\rm Pl}^2}{2}\,
      \frac{(1+2\alpha R)(R-4\Lambda)}{(1+2\alpha R)^3}.
\end{equation}
Using \eqref{eq:dR_dphi}, the first derivative of the potential becomes
\begin{equation}
    V'(\phi)
    = -\frac{\beta M_{\rm Pl}}{2\alpha}\,
      \frac{R-4\Lambda}{(1+2\alpha R)} .
    \label{eq:Vprime}
\end{equation}
Thus $V'(\phi)=0$ precisely at
\begin{equation}
    R = R_d = 4\Lambda ,
\end{equation}
corresponding to the de Sitter background.

\paragraph*{Second derivative.}
Applying the chain rule once more yields
\begin{equation}
    V''(\phi)
    = \frac{\beta^2}{\alpha}\,
      \frac{1-2\alpha R +16\alpha\Lambda}{(1+2\alpha R)^2}.
    \label{eq:Vsecond}
\end{equation}
Evaluating this at the de Sitter point gives
\begin{equation}
    V''(\phi_d)
    = \frac{\beta^2}{\alpha(1+8\alpha\Lambda)} > 0
    \qquad (\alpha>0),
\end{equation}
showing that the de Sitter configuration corresponds to a local minimum of the
Einstein-frame potential and is therefore linearly stable.

\paragraph*{Third derivative.}
For completeness, the third derivative of the potential is
\begin{equation}
    V'''(\phi)
    = \frac{2\beta^3}{\alpha M_{\rm Pl}}\,
      \frac{32\alpha\Lambda - 2\alpha R + 3}{(1+2\alpha R)^2},
    \label{eq:Vthird}
\end{equation}
which is nonvanishing and controls the leading self-interactions of the scalar
mode around the de Sitter minimum.  Equation \eqref{eq:Vthird} is fully
consistent with the explicit expression obtained by direct differentiation in
the Einstein frame.

\paragraph*{Stability interpretation.}
The conditions
\begin{equation}
    V'(\phi_d)=0,
    \qquad
    V''(\phi_d)>0
\end{equation}
establish that the modified Starobinsky model admits a stable de Sitter vacuum solution in metric $f(R)$ gravity. The corresponding scalar degree of freedom has positive mass squared, in agreement with the perturbative analysis of Sec.~\ref{sec:perturbed-ricci}.

The full chameleon mechanism discussed in Sec.~\ref{subsec:chameleon}
requires including the matter coupling through the conformal factor
$A(\phi)$ and analyzing the density dependence of the effective potential $V_{\rm eff}(\phi)=V(\phi)+\rho A(\phi)$. For the present discussion, it is sufficient to note that the vacuum Einstein-frame potential $V(\phi)$ derived from the modified Starobinsky model admits a stable de~Sitter minimum within metric $f(R)$ gravity. The resulting density-dependent scalar mass and screening behavior are encoded in the same Einstein-frame structure already introduced in Sec.~\ref{subsec:chameleon}.

%-------------------------------------------------------------
\subsection{Trace perturbations and the scalar mass}
%-------------------------------------------------------------

We now revisit the trace equation in the Jordan frame in order to extract the explicit mass of the massive scalar propagating mode for our specific $f(R)$ model and to confirm consistency with the general result obtained in Sec.~\ref{sec:perturbed-ricci}.
We decompose the curvature scalar as
\begin{equation}
    R = R_d + \delta R,
\end{equation}
where $R_d$ is the constant-curvature de~Sitter background.

The trace of the vacuum field equations,
\begin{equation}
    3\Box f'(R) + R f'(R) - 2 f(R) = 0,
\end{equation}
may be linearized about the de~Sitter background.
Using $\delta f' = f''(R_d)\,\delta R$ and $\delta f = f'(R_d)\,\delta R$, and retaining terms to first order in $\delta R$, the trace equation reduces to a Klein--Gordon equation for the scalar curvature perturbation,
\begin{equation}
    \Box\,\delta R
    - \frac{1}{3}
      \left(
          \frac{f'(R_d)}{f''(R_d)} - R_d
      \right)\delta R
    = 0.
    \label{KG-trace-general}
\end{equation}
This form makes explicit that $\delta R$ propagates as a massive scalar field, with effective mass
\begin{equation}
    m_\psi^2
    = \frac{1}{3}
      \left(
          \frac{f'(R_d)}{f''(R_d)} - R_d
      \right),
\end{equation}
in agreement with the general expression derived earlier from the linearized field equations.

We now specialize to the modified Starobinsky model,
\begin{equation}
    f(R) = R + \alpha R^2 - 2\Lambda.
\end{equation}
For this choice one finds
\begin{equation}
    f'(R_d) = 1 + 2\alpha R_d,
    \qquad
    f''(R_d) = 2\alpha.
\end{equation}
Substituting these expressions into Eq.~\eqref{KG-trace-general} yields
\begin{equation}
    \left(
        \Box - \frac{1}{6\alpha}
    \right)\delta R = 0,
    \label{KG-eqn-vacuum-massive-scalar-mode}
\end{equation}
so that the scalar curvature perturbation satisfies a Klein--Gordon equation
with mass
\begin{equation}
    m_{\psi}^2 = \frac{1}{6\alpha},
\end{equation}
independent of the background curvature $R_d$.

For the modified Starobinsky model, the de~Sitter background curvature is fixed by the cosmological constant through
\begin{equation}
    R_d = 4\Lambda,
\end{equation}
while the mass of the additional scalar degree of freedom is controlled entirely by the quadratic coupling $\alpha$,
\begin{equation}
    m_{\psi}^{-1} = \sqrt{6\alpha}.
\end{equation}
From a cosmological perspective, $\alpha$ determines the range and dispersion scale of the scalar polarization of gravitational waves, whereas $\Lambda$ fixes the asymptotic de~Sitter curvature. This clean separation of roles will be important when we discuss the
propagation of the scalar mode and its potential observational signatures in de~Sitter cosmology.

\section{SVT Decomposition of the Perturbed Ricci Tensor in Metric \texorpdfstring{$f(R)$}{f(R)} Gravity}

In this section we revisit the $(3+1)$ decomposition in the presence of matter sources. Instead of starting from the vacuum perturbed field equation \eqref{perturbed-field-eqn-vacuum}, we now consider the linearized field equations of the modified Starobinsky model in a nearly Minkowski background, including the stress--energy tensor $T_{\mu\nu}$:
\begin{equation}\label{eqn67}
    \delta G_{\mu\nu}
    + 2\alpha\bigl(\eta_{\mu\nu}\Box-\partial_{\mu}\partial_{\nu}\bigr)\delta R
    = \kappa T_{\mu\nu},
\end{equation}
where $\alpha>0$ is the $R^2$ coupling, $\delta R$ is the scalar curvature perturbation, and $\eta_{\mu\nu}$ is the background Minkowski metric. Using the relation
\begin{equation}
    m_\psi^2 = \frac{1}{6\alpha},
\end{equation}
the term proportional to $\alpha$ can also be written as $(1/3m_\psi^2)(\eta_{\mu\nu}\Box-\partial_\mu\partial_\nu)\delta R$, in agreement with the vacuum analysis.

The Klein--Gordon equation for the massive extra scalar mode \eqref{KG-eqn-vacuum-massive-scalar-mode} in vacuum generalizes in the presence of matter to
\begin{equation}\label{eqn68}
    (\Box-m_{\psi}^2)\,\delta R = m_{\psi}^2\,\kappa\, T,
\end{equation}
or equivalently
\begin{equation}\label{eqn69}
    \Box\delta R = m_{\psi}^2(\delta R+\kappa T),
\end{equation}
where $T \equiv \eta^{\mu\nu}T_{\mu\nu}$ is the trace of the stress--energy tensor. Equation \eqref{eqn68} shows that $\delta R$ behaves as a massive scalar field (the massive scalar propagating mode) sourced by the trace $T$; in the limit $T\to 0$ we recover the vacuum equation.

Using the definition of the Einstein tensor in the flat background,
\begin{equation}
    \delta G_{\mu\nu} = \delta R_{\mu\nu} - \frac{1}{2}\eta_{\mu\nu}\delta R,
\end{equation}
and eliminating $\Box\delta R$ via \eqref{eqn69}, the linearized field equation
\eqref{eqn67} may be written as
\begin{equation}\label{eqn70}
    \delta R_{\mu\nu}
    - \frac{1}{2}\eta_{\mu\nu}\delta R
    + \frac{1}{3}\eta_{\mu\nu}(\delta R + \kappa T)
    - 2\alpha\partial_{\mu}\partial_{\nu}\delta R
    = \kappa T_{\mu\nu},
\end{equation}
or, equivalently,
\begin{equation}\label{eqn71}
    \delta R_{\mu\nu}
    - \biggl(\frac{1}{6}\eta_{\mu\nu}
      + 2\alpha\partial_{\mu}\partial_{\nu}\biggr)\delta R
    = \kappa\Bigl[T_{\mu\nu}-\frac{1}{3}\eta_{\mu\nu}T\Bigr].
\end{equation}
Equations \eqref{eqn70} and \eqref{eqn71} are the starting point for the $(3+1)$ decomposition with matter: the left-hand side contains the usual Ricci-tensor perturbation corrected by the massive scalar propagating mode $\delta R$, while the right-hand side involves the traceless combination $T_{\mu\nu}-\eta_{\mu\nu}T/3$.

%-------------------------------------------------------------
\subsection{Irreducible SVT decomposition of the metric and matter}
%-------------------------------------------------------------

Following \cite{Flanagan_2005,Bardeen_variables}, the SVT decomposition of the metric perturbation $h_{\mu\nu}$ in a nearly Minkowski background reads
\begin{align}
    h_{00} &= 2\psi, \label{eqn72} \\
    h_{0i} &= \beta_i+\partial_i\gamma, \label{eqn73} \\
    h_{ij} &= -2\phi\, \delta_{ij}
      + \Bigl(\partial_i\partial_j-\frac{1}{3}\delta_{ij}\nabla^2\Bigr)\lambda
      + \frac{1}{2}\bigl(\partial_i\epsilon_j+\partial_j\epsilon_i\bigr)
      + h_{ij}^{TT}, \label{eqn74}
\end{align}
where we have defined the new quantities
\(
\psi, \beta_i, \gamma, \phi, \epsilon_i, \lambda, h_{ij}^{TT}
\)
with the assumption that $h_{\mu\nu}\rightarrow0$ as $r\rightarrow\infty$. The transverse and traceless conditions are
\begin{align}
    \partial^i\beta_i &= 0, \label{eqn75}\\
    \partial^i\epsilon_i &= 0, \label{eqn76}\\
    \partial^ih_{ij}^{TT} &= 0, \label{eqn77}\\
    \delta^{ij}h_{ij}^{TT} &= 0. \label{eqn78}
\end{align}

Both in \cite{Flanagan_2005} and in our earlier Bardeen-variable work, it has been shown how the variables
\(
\psi, \beta_i, \gamma, \phi, \epsilon_i, \lambda, h_{ij}^{TT}
\)
transform under a gauge transformation generated by $\xi^a$ with $\xi^a\rightarrow0$ as $r\rightarrow\infty$. Such transformations are parametrized as
\begin{equation}\label{eqn79}
     \xi^a = (\xi^0, \xi^i) = (A, B^i + \partial^i C),
\end{equation}
with $\partial_i B^i=0$. Following the same procedure as in \cite{Flanagan_2005}, one obtains the gauge-invariant scalar and vector combinations
\begin{align}
    \Phi &\equiv -\phi-\frac{1}{6}\nabla^2\lambda, \label{eqn80}\\
    \Psi &\equiv -\psi+\dot\gamma-\frac{1}{2}\ddot\lambda, \label{eqn81}\\
    V_i &\equiv \beta_i-\frac{1}{2}\dot\epsilon_i, \qquad \partial_i V^i=0. \label{eqn82}
\end{align}
The tensor perturbation $h_{ij}^{TT}$ is already gauge invariant.

We can perform a similar SVT decomposition of the matter stress--energy tensor $T_{\mu\nu}$ on the right-hand side of the field equations. We write
\begin{align}
    T_{00} &= \rho, \label{eqn83}\\
    T_{0i} &= S_i+\partial_iS, \label{eqn84}\\
    T_{ij} &= -P\delta_{ij}+\sigma_{ij}
      +(\partial_i\sigma_j+\partial_j\sigma_i)
      +\Bigl(\partial_i\partial_j-\frac{1}{3}\delta_{ij}\nabla^2\Bigr)\sigma,
      \label{eqn85}
\end{align}
where $\rho$, $S$, $S_i$, $P$, $\sigma$, $\sigma_i$, and $\sigma_{ij}$ are new scalar, vector, and tensor quantities with the constraints
\begin{align}
    \partial_iS^i &= 0, \label{eqn86}\\
    \partial^i\sigma_i &= 0, \label{eqn87}\\
    \partial^i\sigma_{ij} &= 0, \label{eqn88}\\
    \delta^{ij}\sigma_{ij} &= 0, \label{eqn89}
\end{align}
along with boundary conditions
\(
S\rightarrow0, \sigma_i\rightarrow0,\sigma\rightarrow0, \nabla^2\sigma\rightarrow0
\)
as $r\rightarrow\infty$ (spatial infinity). The overall minus sign in the isotropic part
$-P\delta_{ij}$ in \eqref{eqn85} will be tracked explicitly in the relations obtained
from stress--energy conservation below.

The conservation law
\begin{equation}\label{eqn90}
    \partial^{\mu}T_{\mu\nu}=0,
\end{equation}
determines relations between $\rho$, $S$, $S_i$, $P$, $\sigma$, $\sigma_i$, and $\sigma_{ij}$.
In the nearly Minkowski background used throughout this section we have
$\partial^{\mu} = (-\partial_0,\partial^i)$, so \eqref{eqn90} reads
\begin{equation}
    -\partial_0 T_{0\nu} + \partial^i T_{i\nu} = 0.
\end{equation}

As a useful special case (and for later physical interpretation), we note that a perfect
fluid at rest in Minkowski space has
\begin{equation}\label{eqn91}
    T_{\mu\nu}=
    \begin{bmatrix}
    \rho &  0 & 0  &  0\\
    0    & -P & 0  &  0\\
    0    &  0 & -P &  0\\
    0    &  0 &  0 & -P\\
    \end{bmatrix},
\end{equation}
so that $T_{0i}=0$ and $T_{ij}=-P\delta_{ij}$. Comparing with \eqref{eqn84}--\eqref{eqn85},
this corresponds to vanishing momentum and anisotropic-stress components
($S_i=0$, $S=0$, $\sigma_i=0$, $\sigma=0$, $\sigma_{ij}=0$), while retaining the scalars
$\rho$ and $P$. In particular, the trace is
\begin{equation}\label{eqn94}
    T=\rho-3P.
\end{equation}

The $\nu=0$ component of \eqref{eqn90} gives
\begin{equation}\label{eqn95}
    \dot\rho=\partial^iT_{i0}=\partial^iT_{0i}=\nabla^2S,
\end{equation}
where we have used \eqref{eqn84} and the constraint \eqref{eqn86}. For $\nu=i$, it is
convenient to separate the two pieces entering
$-\partial_0T_{0i}+\partial^jT_{ji}=0$.
First, taking a spatial derivative of \eqref{eqn85} yields
\begin{equation}\label{eqn96}
    \partial^jT_{ji}
    = -\partial_iP
      +\frac{2}{3}\nabla^2\partial_i\sigma
      +\nabla^2\sigma_i
      +\partial^j\sigma_{ji},
\end{equation}
and the constraints \eqref{eqn87}--\eqref{eqn88} imply $\partial^j\sigma_{ji}=0$.
Second, taking a time derivative of \eqref{eqn84} gives
\begin{equation}\label{eqn97}
    \partial_0T_{0i}=\dot S_i+\partial_i\dot S.
\end{equation}
Stress--energy conservation for $\nu=i$ then combines \eqref{eqn96} and \eqref{eqn97} as 
\begin{equation}\label{eqn99}
    \partial_i\Bigl(-P-\dot S+\frac{2}{3}\nabla^2\sigma\Bigr)
    -\dot S_i+\nabla^2\sigma_i=0,
\end{equation}
where we used $\partial^j\sigma_{ji}=0$.

Taking one more spatial derivative of \eqref{eqn99} and applying the constraints on $\sigma_i$ and $S_i$, we obtain
\begin{equation}\label{eqn100}
    \nabla^2\Bigl[-P-\dot S+\frac{2}{3}\nabla^2\sigma\Bigr]=0.
\end{equation}
Applying the boundary condition at spatial infinity for $S$, $P$, and $\nabla^2\sigma$ (which also guarantees the uniqueness of the decomposition), we conclude that 
\begin{equation}\label{eqn101}
    -P-\dot S+\frac{2}{3}\nabla^2\sigma=0.
\end{equation}
Inserting this condition into \eqref{eqn99} gives
\begin{equation}\label{eqn102}
    \nabla^2\sigma^i=\dot S^i.
\end{equation}
Equation \eqref{eqn101} can also be rewritten as
\begin{equation}\label{eqn103}
    \nabla^2\sigma=\frac{3}{2}\dot S-\frac{3}{2}P.
\end{equation}
Equations \eqref{eqn95}, \eqref{eqn102}, and \eqref{eqn103} are the required set of differential equations that relate the newly defined irreducible matter variables $\rho$, $S_i$, $S$, $P$, $\sigma$, $\sigma_{i}$, and $\sigma_{ij}$. These results match Eq.\ (23) of \cite{MorettiBombacignoMontani2019}, up to differences in notation.

%-------------------------------------------------------------
\subsection{(3+1) Decomposition of the perturbed Ricci tensor}
\label{Perturbed-Ricci-Tensor}
%-------------------------------------------------------------

In $f(R)$ gravity the field equations contain higher-order derivatives of the metric through their dependence on the Ricci scalar $R$. Unlike in General Relativity, where the Einstein tensor $G_{\mu\nu}$ alone determines the dynamics, $f(R)$ theories introduce a massive scalar propagating mode, associated with the scalar curvature perturbation $\delta R$. To fully understand how this scalar mode interacts with the usual scalar, vector, and tensor components of the metric perturbation, it is necessary to go beyond the standard metric decomposition and analyze the perturbation of the Ricci tensor $\delta R_{\mu\nu}$ itself.

By expressing $\delta R_{\mu\nu}$ in terms of the gauge-invariant Bardeen variables and the scalar curvature perturbation, we obtain a set of decoupled differential equations that reveal how each mode behaves in the presence of matter. This decomposition provides a more complete and transparent description of the linearized dynamics in $f(R)$ gravity and is particularly useful for identifying modifications to gravitational-wave propagation and structure formation due to the extra scalar mode.

Following the approach in \cite{Flanagan_2005}, the components of the Einstein tensor $G_{\mu\nu}$ were decomposed in GR to obtain a set of differential equations for the perturbation of the Einstein tensor in terms of the Bardeen variables. In the case of $f(R)$ gravity, we instead decompose the perturbed Ricci tensor $\delta R_{\mu\nu}$ in terms of the Bardeen variables, to obtain a new set of differential equations. The components of $\delta R_{\mu\nu}$ are
\begin{align}
    \delta R_{00} &= \nabla^2\Psi-\frac{3}{2}\ddot\Phi, \label{eqn104}\\
    \delta R_{0i} &= -\frac{1}{2}\nabla^2 V_i-\partial_i\dot\Phi, \label{eqn105}\\
    \delta R_{ij} &= -\partial_{(i}\dot \epsilon_{j)}
      -\partial_i\partial_j\Bigl(\Phi+\frac{1}{2}\Psi\Bigr)
      -\frac{1}{2}\Box h_{ij}^{TT}
      -\delta_{ij}\bigl(-\ddot\Phi+\nabla^2\Phi\bigr).
      \label{eqn106}
\end{align}

In terms of these Bardeen variables $\Phi, \Psi, V_i$, the field equation in the form of Eq.\ \eqref{eqn71} can be recast into a set of differential equations, each corresponding to a component of $\delta R_{\mu\nu}$. For example, the $00$ component of \eqref{eqn71} takes the form
\begin{equation}\label{eqn107}
    \delta R_{00}
    -\frac{1}{6m_{\psi}^2}\bigl(m_{\psi}^2\eta_{00}
    +2\partial_0\partial_0\bigr)\delta R
    =\kappa\Bigl(T_{00}-\frac{1}{3}\eta_{00}T\Bigr).
\end{equation}
Substituting the expression for $\delta R_{00}$ from Eq.\ \eqref{eqn104}, using $\eta_{00}=-1$ and $T=\rho-3P$, we obtain
\begin{equation}\label{eqn108}
    \nabla^2\Psi-\frac{3}{2}\ddot\Phi
    +\frac{1}{6}\delta R
    -\frac{1}{3m_{\psi}^2}\ddot{\delta R}
    =\kappa\biggl(\frac{4}{3}\rho-P\biggr),
\end{equation}
which gives the corresponding differential equation for the $00$ component of $\delta R_{\mu\nu}$.

Next we consider the differential equations corresponding to the $0i$ component of the perturbation of the Ricci tensor. Equation \eqref{eqn71} gives
\begin{equation}\label{eqn109}
    \delta R_{0i}
    -\frac{1}{6m_{\psi}^2}\bigl(m_{\psi}^2\eta_{0i}
    +2\partial_0\partial_i\bigr)\delta R
    =\kappa\Bigl(T_{0i}-\frac{1}{3}\eta_{0i}T\Bigr).
\end{equation}
Substituting the expression \eqref{eqn105} for $\delta R_{0i}$ in Eq.\ \eqref{eqn109}, Eq.\ \eqref{eqn84} for $T_{0i}$, and using $\eta_{0i}=0$, we obtain
\begin{equation}\label{eqn110}
   - \frac{1}{2} \nabla^2 V_i - \partial_i \dot{\Phi}
   - \frac{1}{3m_{\psi}^2} \partial_0\partial_i \delta R
   = \kappa(S_i + \partial_i S).
\end{equation}
At spatial infinity ($r\rightarrow\infty$), we impose $S\rightarrow0$, so that $\partial_i S\rightarrow0$, and similarly $\partial_i\dot\Phi\rightarrow0$ and $\partial_0\partial_i\delta R\rightarrow0$. Under these conditions Eq.\ \eqref{eqn110} reduces to
\begin{equation}\label{eqn111}
    \nabla^2V_i=-2\kappa S_i.
\end{equation}
Separating the longitudinal and transverse parts of \eqref{eqn110} and comparing the coefficients of $\partial_i$ yields
\begin{equation}\label{eqn112}
    \dot\Phi+\frac{1}{3m_{\psi}^2}\dot{\delta R}=-\kappa S.
\end{equation}
Equations \eqref{eqn111} and \eqref{eqn112} are the differential equations based on the $0i$ component of $\delta R_{\mu\nu}$ in terms of the Bardeen variables and the massive scalar propagating mode.

Finally, we consider the $ij$ component of the perturbation of the Ricci tensor, which is
\begin{equation}\label{eqn113}
     \delta R_{ij}
     -\frac{1}{6m_{\psi}^2}\bigl(m_{\psi}^2\eta_{ij}
     +2\partial_i\partial_j\bigr)\delta R
     =\kappa\Bigl(T_{ij}-\frac{1}{3}\eta_{ij}T\Bigr).
\end{equation}
Substituting the expression \eqref{eqn106} for $\delta R_{ij}$ and Eq.\ \eqref{eqn85} for $T_{ij}$ into Eq.\ \eqref{eqn113}, and equating coefficients of the independent SVT pieces, we obtain
\begin{align}
    -\partial_{(i}\dot \epsilon_{j)} &= \kappa\partial_{(i}\sigma_{j)}, \label{eqn114}\\
    -\frac{1}{2}\Box h_{ij}^{TT} &= \kappa\sigma_{ij}, \label{eqn115}\\
    -\frac{1}{2}\Box\delta_{ij}\Phi
    -\frac{1}{3}\partial_i\partial_j\delta R
    &=\kappa\Bigl(-\frac{1}{3}\delta_{ij}\nabla^2\sigma
       -\frac{1}{3}\eta_{ij}T\Bigr). \label{eqn116}
\end{align}
These equations imply
\begin{align}
    \dot V_i &= \kappa\sigma_i, \label{eqn117}\\
    \Box h_{ij}^{TT} &= -2\kappa\sigma_{ij}, \label{eqn118}\\
    \Psi+\frac{1}{2}\Phi +\frac{1}{3m_{\psi}^2}\delta R &= -\kappa\sigma,
      \label{eqn119}
\end{align}
and, using $T=\rho-3P$,
\begin{equation}\label{eqn120}
    \Box\Phi+\frac{1}{3}\delta R
    =\frac{2}{3}\kappa\bigl(\nabla^2\sigma-\rho\bigr).
\end{equation}
Equations \eqref{eqn117}–\eqref{eqn120} are the set of differential equations corresponding to the $ij$ component of $\delta R_{\mu\nu}$ in terms of the Bardeen variables and the matter SVT variables. These results are consistent with those derived in \cite{MorettiBombacignoMontani2019}, up to differences in notation.

%-------------------------------------------------------------
\subsection{Cosmological interpretation of the SVT equations with the extra scalar degree of freedom}
%-------------------------------------------------------------

The system of equations \eqref{eqn108}, \eqref{eqn111}, \eqref{eqn112}, and \eqref{eqn117}–\eqref{eqn120} allows a direct physical interpretation in cosmology once the background is promoted from Minkowski to a slowly varying FLRW or de Sitter spacetime.

The $\delta R_{00}$ equation \eqref{eqn108} is a modified Poisson-type equation: the gravitational potential $\Psi$ is sourced not only by the energy density $\rho$, but also by pressure $P$, time derivatives of the potential $\Phi$, and the dynamics of the scalar curvature perturbation $\delta R$ \cite{Tsujikawa_2007,Song_2007}. In GR, the corresponding equation at linear order would involve essentially the Laplacian of $\Psi$ sourced only by $\rho$, with no extra scalar degree of freedom contribution. This modification leads to a scale- and time-dependent effective gravitational coupling, which directly affects the growth of cosmological structure and can be constrained by large-scale structure and weak-lensing surveys.

The $\delta R_{0i}$ sector separates into a transverse (vector) part and a longitudinal (scalar) part. The transverse part, Eq.\ \eqref{eqn111} together with Eq.\ \eqref{eqn117}, shows that vector perturbations ${V_i}$ are sourced by the transverse momentum density $S_i$ and anisotropic stress $\sigma_i$, just as in GR. Thus $f(R)$ gravity does not introduce new propagating vector modes at linear order. The longitudinal scalar equation \eqref{eqn112}, however, contains the time derivative of the additional scalar degree of freedom $\dot{\delta R}$, modifying the time evolution of $\Phi$ relative to GR. The time dependence of the gravitational potentials is directly probed by the integrated Sachs–Wolfe (ISW) effect and cross-correlations of CMB maps with large-scale structure.

The $\delta R_{ij}$ equations show that the tensor sector, Eq.\ \eqref{eqn118}, obeys a wave equation structurally identical to that of GR, but with a source from anisotropic stress. In $f(R)$ gravity, the background scalar degree of freedom and the modified expansion history can nevertheless change the amplitude damping and effective propagation of gravitational waves over cosmological distances, providing an additional channel to test modifications of gravity with standard sirens.

Finally, the scalar sector of the perturbed field equations provides a direct window into one of the characteristic phenomenological signatures of modified gravity. In linear cosmological perturbation theory, scalar metric perturbations are described by the gauge-invariant Bardeen potentials $\Phi$ and $\Psi$, which coincide in General Relativity in the absence of
matter anisotropic stress. Their inequality, $\Phi \neq \Psi$, is commonly referred to as \emph{gravitational slip} and signals a departure from GR caused either by imperfect fluids or by additional gravitational degrees of freedom
\cite{Song_2007,Bean_2010}.

In metric $f(R)$ gravity, the scalar part of the $\delta R_{ij}$ equations, Eqs.~\eqref{eqn119} and \eqref{eqn120}, reveals that gravitational slip arises generically even when the matter anisotropic stress vanishes ($\sigma=0$, equivalently $\Pi=0$). In this case, the difference between the two scalar potentials is instead sourced by the scalar curvature perturbation
$\delta R$, reflecting the presence of the propagating scalar mode. This modification of the relation between $\Phi$ and $\Psi$ is a robust signature of $f(R)$ models and can be observationally constrained through joint analyses of galaxy clustering, redshift-space distortions, and weak gravitational lensing \cite{Song_2007,Xu_2015}. The full SVT decomposition of
$\delta R_{\mu\nu}$ thus provides a unified framework for linking the gauge-invariant scalar dynamics of the theory to observable effects in both gravitational-wave physics and cosmology.

\section{Geodesic deviation method to find the polarization content}
\label{sec:geodev}

The geodesic deviation equation relates the Riemann curvature tensor to the relative acceleration of neighboring geodesics and therefore provides a direct probe of gravitational-wave polarizations in a given theory of gravity
\cite{Carroll:2004st,MisnerThorneWheeler1973}. In this section we use the geodesic deviation equations to identify the polarization modes of gravitational waves in our specific metric $f(R)$ model,
\begin{equation}
  f(R) = R + \alpha R^2 - 2\Lambda,
\end{equation}
for which the scalar curvature perturbation $\delta R$ obeys the massive Klein--Gordon equation
\begin{equation}
  \bigl(\Box - m_\psi^2\bigr)\,\delta R = 0,
  \qquad
  m_\psi^2 = \frac{1}{6\alpha},
\end{equation}
on a de Sitter background. The scalar perturbation $\delta R$ corresponds to the extra scalar degree of freedom, in addition to the usual tensor modes of GR.

We first work in the local Minkowski patch of the de Sitter background, which is appropriate for interferometric detectors whose size is much smaller than the background curvature radius. We then show how the same polarization structure appears when the calculation is formulated fully on a de Sitter FRW background.

%-------------------------------------------------------------
\subsection{Local Minkowski patch of de Sitter}
\label{subsec:geodev-minkowski}
%-------------------------------------------------------------

The general geodesic deviation equation is
\begin{equation}\label{eqn121}
    \frac{D^2\xi^{\mu}}{d\tau^2}
    = -R^{\mu}{}_{\alpha\nu\beta}\,
      \xi^{\nu}
      \frac{dx^{\alpha}}{d\tau}
      \frac{dx^{\beta}}{d\tau},
\end{equation}
where $\xi^\mu$ is the separation vector between neighboring geodesics and $\tau$ is proper time. For gravitational-wave detectors we work in the weak-field, slow-motion limit: the detector is at rest in the chosen coordinates and far from the source, so
\begin{equation}\label{eqn122}
  \frac{dx^i}{d\tau} \ll \frac{dx^0}{d\tau}
  \quad\Rightarrow\quad
  \frac{dx^\mu}{d\tau} \approx (1,0,0,0),
\end{equation}
and we can identify proper time with coordinate time,
\begin{equation}\label{eqn123}
  \tau \simeq t.
\end{equation}
In this regime the covariant derivatives in \eqref{eqn121} reduce to ordinary time derivatives, and the spatial components of the geodesic deviation equation become
\begin{equation}\label{eqn124}
  \ddot{\xi}_i = -R_{i0j0}\,\xi^j,
\end{equation}
where overdots denote derivatives with respect to $t$.  

In linearized gravity, the Riemann tensor is
\begin{equation}
  R_{\mu\nu\rho\sigma}
  = \frac{1}{2}\Bigl(
      \partial_{\rho}\partial_{\nu} h_{\mu\sigma}
      + \partial_{\sigma}\partial_{\mu} h_{\nu\rho}
      - \partial_{\sigma}\partial_{\nu} h_{\mu\rho}
      - \partial_{\rho}\partial_{\mu} h_{\nu\sigma}
    \Bigr),
\end{equation}
where $h_{\mu\nu}$ is the metric perturbation on the local Minkowski background $\eta_{\mu\nu}$. We decompose $h_{\mu\nu}$ into a transverse-traceless tensor part $h_{\mu\nu}^{TT}$ and a scalar part associated with the scalar curvature perturbation $\delta R$.

For the scalar mode, in a convenient gauge compatible with the Newtonian (longitudinal) gauge used in Section~\ref{sec:3p1}, the scalar perturbation can be chosen proportional to the background metric:
\begin{equation}
  h^{(s)}_{\mu\nu} = C\,\delta R\, \eta_{\mu\nu},
\end{equation}
where $C$ is an overall constant that only rescales the amplitude and does not affect the polarization pattern. For simplicity we set $C=1$ below.

Explicitly,
\begin{align}
  h^{(s)}_{00} &= -\delta R, \\
  h^{(s)}_{0i} &= 0, \\
  h^{(s)}_{ij} &= \delta R\,\delta_{ij}.
\end{align}

Substituting into the expression for the Riemann tensor and focusing on $R_{i0j0}$, we find
\begin{equation}\label{Riemann-component-eqn}
  R_{i0j0}
  = -\frac{1}{2}\bigl(
      \partial_0\partial_0 h_{ij}
      + \partial_i\partial_j h_{00}
    \bigr)
  = -\frac{1}{2}\Bigl(
      \delta_{ij}\,\ddot{\delta R}
      - \partial_i\partial_j \delta R
    \Bigr).
\end{equation}

Now consider a scalar wave propagating along the $+z$ direction,
\begin{equation}
  \delta R = \delta R(t,z).
\end{equation}
In the transverse directions $x$ and $y$,
\begin{equation}
  \partial_x \delta R = \partial_y \delta R = 0,
\end{equation}
so that
\begin{equation}
  R_{x0x0} = R_{y0y0}
  = -\frac{1}{2}\ddot{\delta R}.
\end{equation}
In the longitudinal direction,
\begin{equation}
  R_{z0z0}
  = -\frac{1}{2}\Bigl(
      \ddot{\delta R}
      - \partial_z^2 \delta R
    \Bigr).
\end{equation}

Using the massive Klein--Gordon equation for the extra scalar mode in the local Minkowski patch,
\begin{equation}\label{eqn132}
  \bigl(\Box - m_\psi^2\bigr)\,\delta R
  = 0,
\end{equation}
which implies
\begin{equation}
  \ddot{\delta R} = \partial_z^2 \delta R - m_\psi^2 \delta R,
\end{equation}
we obtain
\begin{align}
  R_{x0x0} &= R_{y0y0}
    = -\frac{1}{2}\bigl(\partial_z^2 \delta R - m_\psi^2 \delta R\bigr), \\
  R_{z0z0}
    &= +\frac{1}{2} m_\psi^2 \delta R.
\end{align}

For a monochromatic plane wave
\begin{equation}
  \delta R(t,z) = A\,e^{i(k_z z - \omega t)},
  \qquad
  \omega^2 = k_z^2 + m_\psi^2,
\end{equation}
the tidal components become
\begin{align}
  R_{x0x0} &= R_{y0y0}
    = -\frac{1}{2}\omega^2\,\delta R, \\
  R_{z0z0}
    &= +\frac{1}{2} m_\psi^2 \,\delta R.
\end{align}

The geodesic deviation equations
\begin{equation}
  \ddot{\xi}_i = - R_{i0j0}\,\xi^j
\end{equation}
then give 
\begin{align}
  \ddot{X}
    &= -\frac{1}{2}\omega^2\,\delta R\,X, \\
  \ddot{Y}
    &= -\frac{1}{2}\omega^2\,\delta R\,Y, \\
  \ddot{Z}
    &= -\frac{1}{2}m_\psi^2\,\delta R\,Z.
\end{align}

These equations show that the extra scalar degree of freedom induces both a transverse breathing mode (in the $X$ and $Y$ directions) and a longitudinal mode (in the $Z$ direction). This is precisely the expected polarization content for a massive scalar mode.

In pure GR, where only the transverse-traceless tensor $h_{ij}^{TT}$ is present, $\delta R = 0$ and the scalar-induced contributions vanish; only the familiar $\oplus$ and $\otimes$ tensor modes remain. In metric $f(R)$ gravity, the nonzero $\delta R$ generates additional breathing and longitudinal polarizations on top of the GR tensor modes.

%-------------------------------------------------------------
\subsection{Geodesic deviation on a de Sitter FRW background}
\label{subsec:geodev-dS}
%-------------------------------------------------------------

We now sketch how the same polarization structure arises when the calculation is performed
directly on the de~Sitter background without passing explicitly to a Minkowski patch. In
spatially flat FRW coordinates, the de~Sitter metric can be written as
\begin{equation}
  ds^2 = -dt^2 + a^2(t)\,\delta_{ij}dx^i dx^j,
  \qquad
  a(t) = e^{H_d t},
\end{equation}
with constant Hubble parameter $H_d = \sqrt{R_d/12}$ in four dimensions.

We consider small perturbations around this background in Newtonian gauge. Restricting
initially to the scalar sector, the perturbed metric takes the form 
\begin{equation}\label{FRW perturbation}
  ds^2 = -(1+2\Phi)\,dt^2
         + a^2(t)(1-2\Psi)\,\delta_{ij}dx^i dx^j,
\end{equation}
where equality of the Bardeen potentials $\Phi=\Psi$ holds for metric $f(R)$ gravity on a de~Sitter background (Section~\ref{sec:3p1}), with
\begin{equation}
  \Phi = \Psi = -\frac{\delta R}{6 m_\psi^2}.
\end{equation}
Thus the additional scalar degree of freedom is directly encoded in both the temporal and isotropic spatial perturbations of the metric.

To relate the geodesic deviation equation to the detector frame, it is convenient to introduce an orthonormal tetrad adapted to a comoving observer,
\begin{align}
  e_{\hat{0}}{}^\mu &= (1,0,0,0), \\
  e_{\hat{i}}{}^\mu &= \frac{1}{a(t)}\,\delta_i{}^\mu,
\end{align}
so that physical (proper) spatial separations are measured with hatted indices. In this
orthonormal frame the geodesic deviation equation takes the form
\begin{equation}
  \frac{d^2\xi^{\hat{i}}}{dt^2}
  = - R^{\hat{i}}{}_{\hat{0}\hat{j}\hat{0}}\,\xi^{\hat{j}} .
\end{equation}

The tidal tensor splits naturally into a background de~Sitter contribution and a
perturbation induced by scalar and tensor modes,
\begin{equation}
  R^{\hat{i}}{}_{\hat{0}\hat{j}\hat{0}}
  = R^{\hat{i}}{}_{\hat{0}\hat{j}\hat{0}}\big|_{\rm dS}
    + \delta R^{\hat{i}}{}_{\hat{0}\hat{j}\hat{0}} .
\end{equation}
For the spatially flat de~Sitter background, the nonvanishing Christoffel symbols are
\(
\Gamma^{0}{}_{ij} = a\dot a\,\delta_{ij}
\)
and
\(
\Gamma^{i}{}_{0j} = H_d\,\delta^{i}{}_{j},
\)
which follow directly from the FRW line element. From these, the coordinate-basis Riemann component relevant for geodesic deviation is
\begin{equation}
  R^{i}{}_{0j0} = \frac{\ddot a}{a}\,\delta^{i}{}_{j}.
\end{equation}
For exact de~Sitter expansion $a(t)=e^{H_d t}$, one has $\ddot a/a=H_d^2$, yielding
\begin{equation}
  R^{i}{}_{0j0}\big|_{\rm dS}=H_d^2\,\delta^{i}{}_{j}.
\end{equation}
Projecting onto the orthonormal tetrad,
\(
R^{\hat{i}}{}_{\hat{0}\hat{j}\hat{0}}
= e^{\hat{i}}{}_{k}\,e_{\hat{j}}{}^{\ell}\,R^{k}{}_{0\ell 0},
\)
the factors of $a(t)$ from the tetrads cancel those implicit in the metric, leaving
\begin{equation}
  R^{\hat{i}}{}_{\hat{0}\hat{j}\hat{0}}\big|_{\rm dS}
  = H_d^2\,\delta^{i}{}_{j}.
\end{equation}

We now include perturbations. Restoring both scalar and tensor modes, the perturbed FRW
metric may be written as 
\begin{equation}
  ds^2 = -(1+2\Phi)\,dt^2
         + a^2(t)\Bigl[(1-2\Psi)\,\delta_{ij}
         + h^{\rm TT}_{ij}(t,\mathbf{x})\Bigr]dx^i dx^j ,
\end{equation}
where $h^{\rm TT}_{ij}$ denotes the transverse--traceless tensor perturbation. Introducing
the Minkowski metric $\eta_{\mu\nu}=\mathrm{diag}(-1,1,1,1)$, all perturbations can be
collected into a single tensor
\begin{equation}
  h_{00}=2\Phi, \qquad
  h_{0i}=0, \qquad
  h_{ij}=2\Psi\,\delta_{ij}+h^{\rm TT}_{ij},
\end{equation}
so that the metric assumes the conformal form
\begin{equation}
  g_{\mu\nu}(t,\mathbf{x})
  = a^2(t)\,\bigl[\eta_{\mu\nu}+h_{\mu\nu}(t,\mathbf{x})\bigr].
\end{equation}

Expanding the Riemann tensor to first order,
\(
R_{\mu\nu\rho\sigma}=R^{(0)}_{\mu\nu\rho\sigma}
+\delta R_{\mu\nu\rho\sigma}[h],
\)
the linearized part depends only on derivatives of $h_{\mu\nu}$. Because the conformal factor
multiplies both background and perturbation, one finds
\begin{equation}
  \delta R_{\mu\nu\rho\sigma}[g]
  = a^2(t)\,\delta R_{\mu\nu\rho\sigma}^{\rm (Mink)}[h].
\end{equation}
Raising indices and projecting onto the orthonormal tetrad yields
\begin{equation}
  \delta R^{\hat{i}}{}_{\hat{0}\hat{j}\hat{0}}
  = \frac{1}{a^2(t)}\,\delta R_{i0j0}^{\rm (Mink)},
\end{equation}
where $\delta R_{i0j0}^{\rm (Mink)}$ is precisely the tidal matrix obtained in
Subsection~\ref{subsec:geodev-minkowski}.

Therefore, for the massive scalar propagating mode, we can directly carry over the Minkowski result, with careful attention to the overall sign,
\begin{align}
  \delta R^{\hat{x}}{}_{\hat{0}\hat{x}\hat{0}}
  &= \delta R^{\hat{y}}{}_{\hat{0}\hat{y}\hat{0}}
   = -\frac{1}{2a^2}\,\omega^2\,\delta R, \\
  \delta R^{\hat{z}}{}_{\hat{0}\hat{z}\hat{0}}
  &= +\frac{1}{2a^2}\,m_\psi^2\,\delta R,
\end{align}
for a monochromatic mode, and similarly for a generic wave packet using the Klein--Gordon
equation \eqref{eqn132}. The overall factor $1/a^2(t)$ dilutes the tidal amplitude due to
cosmic expansion, while leaving the polarization pattern unchanged.

The geodesic deviation equations for physical separations $\xi^{\hat{i}}$ are therefore
\begin{align}
  \ddot{\xi}^{\hat{X}}
    &= - \bigl[
          H_d^2
          + \delta R^{\hat{x}}{}_{\hat{0}\hat{x}\hat{0}}
        \bigr]\xi^{\hat{X}}, \\
  \ddot{\xi}^{\hat{Y}}
    &= - \bigl[
          H_d^2
          + \delta R^{\hat{y}}{}_{\hat{0}\hat{y}\hat{0}}
        \bigr]\xi^{\hat{Y}}, \\
  \ddot{\xi}^{\hat{Z}}
    &= - \bigl[
          H_d^2
          + \delta R^{\hat{z}}{}_{\hat{0}\hat{z}\hat{0}}
        \bigr]\xi^{\hat{Z}}.
\end{align}
The background term produces the isotropic de~Sitter expansion, while the wave-induced part
reproduces the same transverse breathing and longitudinal pattern as in the local Minkowski
analysis. Thus, cosmological expansion modifies amplitudes but does not change the
polarization content.

\subsection{Polarization classification via \texorpdfstring{$R_{i0j0}$}{Ri0j0}}
\label{subsec:polarization-matrix}
%-------------------------------------------------------------

The geodesic deviation equations derived above show explicitly that the additional scalar
degree of freedom in our $f(R)$ model produces both breathing and longitudinal motion of test
particles. For completeness, we now review a more formal method to classify the polarization
modes using the components of $R_{i0j0}$, following
\cite{Eardley,Dong_2024}.

In a local inertial (Minkowski) patch of the spacetime, the perturbed metric may be written in
terms of scalar, vector, and tensor perturbations as
\begin{equation}
  ds^2 = -(1+2\Phi)\,dt^2
         + 2E_i\,dt\,dx^i
         + \bigl[(1-2\Psi)\delta_{ij}
         + h^{TT}_{ij}\bigr]dx^i dx^j ,
\end{equation}
where $\Phi$ and $\Psi$ are the scalar Bardeen potentials (with $\Phi=\Psi$ in the present
context), $E_i$ encodes the vector (shear) perturbations, and $h^{TT}_{ij}$ is the
transverse--traceless tensor mode. To linear order in the perturbations, the Riemann tensor
components entering the geodesic deviation equation are
\begin{equation}\label{eq:Riemann-i0j0}
  R_{i0j0}
  = \partial_i\partial_j\Psi
    - \frac{1}{2}\delta_{ij}\,\partial_0^2\Phi
    - \frac{1}{2}\bigl(\partial_0\partial_i E_j
                      + \partial_0\partial_j E_i\bigr)
    - \frac{1}{2}\partial_0^2 h^{TT}_{ij}.
\end{equation}

\begin{figure}[htp]
    \centering
    \includegraphics{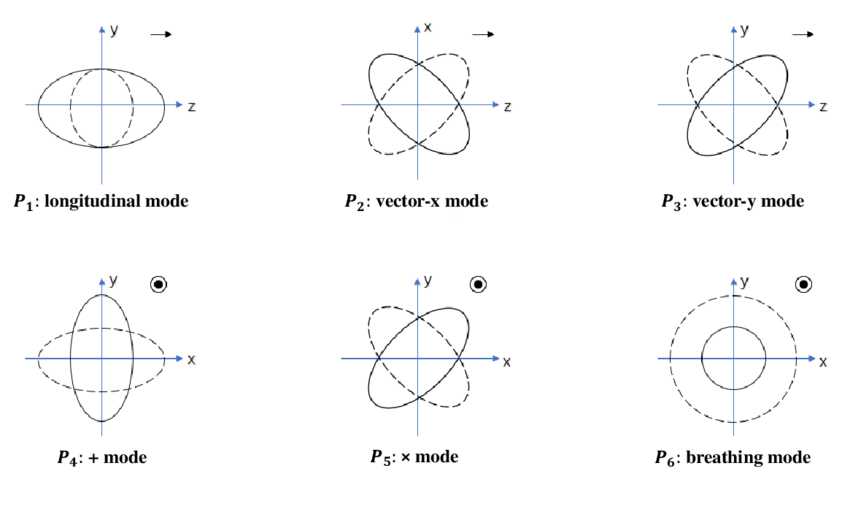}
    \caption{Six polarization modes of gravitational waves.}
    \label{fig:polarization modes}
\end{figure}

The six possible GW polarization modes can be encoded by writing the tidal tensor
$R_{i0j0}$ as a symmetric $3\times 3$ matrix,
\begin{equation}\label{Riemann-matrix}
    R_{i0j0}=
    \begin{bmatrix}
        P_4 + P_6  &  P_5      &  P_2\\
        P_5        & -P_4+P_6  &  P_3\\
        P_2        &  P_3      &  P_1
    \end{bmatrix},
\end{equation}
where $P_1,\dots,P_6$ are the six independent polarization amplitudes (scalar longitudinal,
two vector modes, two tensor modes, and scalar breathing). They correspond to the six standard
polarization patterns shown in Fig.~\ref{fig:polarization modes}.\footnote{This image ``Six
polarization modes of gravitational waves'' is reproduced from \cite{Dong_2024}, and is
licensed under Creative Commons Attribution 4.0 International
(https://creativecommons.org/licenses/by/4.0/).}

For a plane wave propagating along the $z$ direction, comparison of
\eqref{eq:Riemann-i0j0} with the matrix form \eqref{Riemann-matrix} yields
\begin{align}
    P_1 &= \partial_3\partial_3\Psi
          -\frac{1}{2}\partial_0\partial_0\Phi, \label{P1}\\
    P_2 &= \frac{1}{2}\partial_0\partial_3 E_1, \label{P2}\\
    P_3 &= \frac{1}{2}\partial_0\partial_3 E_2, \label{P3}\\
    P_4 &= -\frac{1}{2}\partial_0\partial_0 h_{11}^{TT}, \label{P4}\\
    P_5 &= -\frac{1}{2}\partial_0\partial_0 h_{12}^{TT}, \label{P5}\\
    P_6 &= -\frac{1}{2}\partial_0\partial_0\Phi. \label{P6}
\end{align}
Here $E_1$ and $E_2$ encode the vector (shear) polarizations, $h^{TT}_{ij}$ represents the
usual $\oplus$ and $\otimes$ tensor modes, and $\Phi,\Psi$ are the scalar Bardeen potentials.

In metric $f(R)$ gravity we have generic vector perturbations $V_i = 0$  in vacuum, so the vector modes are absent and $P_2 = P_3 = 0$. The tensor modes $P_4$ and $P_5$ coincide with those of GR and correspond to the $\oplus$ and $\otimes$ polarizations. The remaining scalar modes are encoded in $P_1$ (longitudinal mode, involving both $\Phi$ and $\Psi$) and $P_6$ (breathing mode, involving only $\Phi$). Because $\Phi=\Psi\neq 0$ in our model and are related to the additional scalar degree of freedom via
\begin{equation}
  \Phi = \Psi = -\frac{\delta R}{6m_\psi^2},
\end{equation}
both $P_1$ and $P_6$ are nonzero, confirming that the model exhibits a mixed longitudinal and
breathing scalar polarization in addition to the two tensor polarizations.

In summary, the geodesic deviation analysis---both in the local Minkowski patch and on the
full de~Sitter background---shows that the metric $f(R)$ model
$f(R)=R+\alpha R^2-2\Lambda$ supports:
\begin{itemize}
  \item two massless tensor modes $(\oplus,\otimes)$, identical to those of GR;
  \item one massive scalar mode (the massive scalar propagating mode), which decomposes into a transverse breathing polarization and a longitudinal polarization along the propagation direction.
\end{itemize}
This pattern agrees with the general expectation for metric $f(R)$ gravity and provides the polarization content against which current and future GW observations can test this class of models.

%%%%%%%%%%%%%%%%%%%%%%%%%%%%%%%%%%%%%%%%%%%%%%%%%%%%%%%%%%%%%%%
\section{Conclusion and Future Outlook}
%%%%%%%%%%%%%%%%%%%%%%%%%%%%%%%%%%%%%%%%%%%%%%%%%%%%%%%%%%%%%%%

In this work we developed a unified and fully gauge-invariant analysis of gravitational-wave polarizations in metric $f(R)$ gravity, with particular emphasis on the modified Starobinsky model
\(
    f(R)=R+\alpha R^{2}-2\Lambda .
\)
Working on a constant-curvature de Sitter background, we reformulated the linearized field equations in terms of Bardeen gauge-invariant variables and the scalar curvature perturbation $\delta R$, thereby making the massive scalar propagating mode manifest. By deriving the Klein--Gordon equation for $\delta R$ directly from the perturbed trace equation, we verified that the scalar mode behaves as a massive propagating field with mass $m_\psi^2 = 1/(6\alpha)$ on the de Sitter background. This establishes the scalar curvature perturbation $\delta R$ as the source of the additional breathing and longitudinal polarizations absent in General Relativity.

We complemented the Bardeen-variable analysis with a full $(3+1)$ decomposition of the perturbed Ricci tensor, including the presence of matter sources. This approach revealed explicitly how scalar, vector, and tensor perturbations enter the modified field equations and how the scalar sector departs from its
GR behavior. In particular, the decomposition demonstrated that (i) the vector sector remains nondynamical and identical to that of GR, (ii) the tensor sector continues to satisfy the standard transverse--traceless wave equation, and (iii) all modifications are encoded in the scalar sector through the dynamical
curvature perturbation $\delta R$. The resulting coupled equations for $\Phi$, $\Psi$, and $\delta R$ illustrate the origin of the gravitational slip, modified Poisson equation, and scale-dependent evolution of cosmological perturbations characteristic of $f(R)$ models.

A complementary geodesic-deviation analysis was carried out in both the local-Minkowski patch of de Sitter spacetime and in the fully covariant de Sitter background. In both cases, the tidal tensor $R_{i0j0}$ depends on the scalar curvature perturbation $\delta R$ and yields the characteristic polarization pattern: two tensor modes ($\oplus$ and $\otimes$), a breathing mode, and a longitudinal mode. This
agrees with the general classification of metric theories admitting up to six polarizations and verifies, by two independent methods, that metric $f(R)$ gravity predicts exactly three observable polarization sectors: two tensor and one massive scalar.

From a cosmological perspective, the mass $m_\psi$ of the massive scalar propagating mode sets the transition scale between GR-like behavior at high wavenumbers and modified gravity effects on large scales. Because the same massive scalar propagating mode controls the late-time background evolution, the growth rate of structure, and the propagation of gravitational waves, future multi-probe observations---combining large-scale structure, weak lensing, CMB anisotropies, pulsar-timing arrays, and gravitational-wave observatories---provide a coherent program for testing the viability of $f(R)$ gravity on both astrophysical and cosmological scales.

\subsection*{Future Outlook}

Several natural extensions follow from the framework developed here:

\begin{itemize}
    \item \textbf{Beyond de Sitter backgrounds:}  
    The methods employed here can be generalized to slowly evolving FLRW backgrounds, permitting a direct link between gravitational-wave propagation and the time dependence of the mass $m_{\psi}$ of the massive scalar propagating mode in realistic cosmologies.

    \item \textbf{Mode mixing and GW propagation:}  
    A next step is the study of mode mixing between the tensor and scalar sectors, including amplitude damping and potential dispersion effects in late-time, low-density environments.

    \item \textbf{Constraints from forthcoming surveys:}  
    Current and future missions (Euclid, LSST, SKA, LISA, pulsar-timing arrays) will significantly improve constraints on gravitational slip, the mass of the scalar mode, and the scale-dependent growth of cosmological perturbations. The gauge-invariant formalism presented here is well suited for connecting theoretical predictions with these upcoming datasets.

    \item \textbf{Extension to broader modified-gravity families:}  
    The techniques developed in this paper---decomposition of the perturbed Ricci tensor, isolation of the massive scalar propagating mode, fully covariant GW polarization extraction, and geodesic-deviation analysis---can be applied to more general higher-curvature theories such as $f(G)$ gravity, scalar--tensor Horndeski theories, and Einstein--dilaton--Gauss--Bonnet models.
\end{itemize}

Overall, the combination of gauge-invariant SVT analysis, Ricci-tensor decomposition, and geodesic deviation provides a robust framework for identifying and interpreting the polarization content of gravitational waves in metric $f(R)$ gravity. This establishes a consistent pathway for future observational tests capable of distinguishing GR from its simplest and most theoretically motivated extensions.

\section*{Appendix A: d'Alembertian in Curved Spacetime}

The d'Alembertian operator acting on a scalar field \( \varphi \) in curved spacetime is defined as
\begin{equation}
    \Box \varphi
    = \frac{1}{\sqrt{-g}} \partial_\mu \left( \sqrt{-g}\, g^{\mu\nu} \partial_\nu \varphi \right) 
    = g^{\mu\nu} \nabla_\mu \nabla_\nu \varphi .
\end{equation}

We consider the de Sitter spacetime written in spatially flat Friedmann--Robertson--Walker (FRW) coordinates:
\begin{equation}
    ds^2 = -dt^2 + a^2(t) (dx^2 + dy^2 + dz^2),
\end{equation}
where the scale factor is \( a(t) = e^{Ht} \) and \( H \) is the constant Hubble parameter.

The determinant of the metric is
\begin{equation}
    g = \det(g_{\mu\nu}) = -a^6(t),
\end{equation}
so that
\begin{equation}
    \sqrt{-g} = a^3(t).
\end{equation}

Substituting these into the definition of \( \Box \), we obtain
\begin{align}
    \Box \varphi
    &= \frac{1}{a^3(t)} 
       \partial_\mu \left[ a^3(t)\, g^{\mu\nu} \partial_\nu \varphi \right] \\
    &= \frac{1}{a^3(t)} 
       \left[ \partial_t \left( a^3 g^{tt} \partial_t \varphi \right)
            + \partial_i \left( a^3 g^{ij} \partial_j \varphi \right)
       \right].
\end{align}

Since \( g^{tt} = -1 \) and \( g^{ij} = a^{-2}(t) \delta^{ij} \), and because \( a(t) \) depends only on \( t \), the spatial derivatives of \( a^3(t) \) vanish. Thus,
\begin{align}
    \Box \varphi
    &= -\frac{1}{a^3} \partial_t \left( a^3 \partial_t \varphi \right)
       + \frac{1}{a^3} \partial_i \left( a^3 a^{-2} \delta^{ij} \partial_j \varphi \right) \\
    &= -\frac{1}{a^3} \left( a^3 \partial_t^2 \varphi + 3a^2 \dot{a} \partial_t \varphi \right)
       + \frac{1}{a^2} \nabla^2 \varphi \\
    &= -\partial_t^2 \varphi
       - 3 \frac{\dot{a}}{a}\, \partial_t \varphi
       + \frac{1}{a^2} \nabla^2 \varphi .
\end{align}

For de Sitter spacetime, \( a(t) = e^{Ht} \), so \( \dot{a}/a = H \). Therefore, the d'Alembertian simplifies to
\begin{equation}
    \boxed{
        \Box \varphi
        = -\partial_t^2 \varphi
          - 3H\, \partial_t \varphi
          + e^{-2Ht}\, \nabla^2 \varphi
    }
\end{equation}
where \( \nabla^2 = \partial_x^2 + \partial_y^2 + \partial_z^2 \) is the flat-space Laplacian.

This is the standard expression for the action of the d'Alembertian on a scalar field in a spatially flat de Sitter spacetime written in FRW coordinates.

\bibliographystyle{unsrt}  % or plainnat / unsrtnat depending on template
\bibliography{refs}       % your file should be refs.bib

\end{document}